\documentclass[structabstract]{raa_twocloumn}
\usepackage{graphicx,times}
\usepackage{threeparttable}
\usepackage{natbib,amssymb}
\usepackage{float}
\usepackage{color}
\usepackage{bm}
\usepackage{subfigure}
\usepackage[figuresright]{rotating}
\usepackage{amsmath}
\usepackage{multirow}
\usepackage{url}
\usepackage[backref]{hyperref}

\usepackage{amssymb}
\usepackage{amsmath}
\usepackage{lipsum}
\usepackage{rotating}
\usepackage{comment}

\DeclareUnicodeCharacter{2212}{\textendash}

\graphicspath{{./}{figures/}}
\begin{document}
   \title{Hubble WFC3 Spectroscopy of the Terrestrial Planets L~98-59~c \& d: No Evidence for a Clear Hydrogen Dominated Primary Atmosphere}
   \volnopage{ {\bf 2022} Vol.\ {\bf XX} No. {\bf XXX}, 000--000}
   \setcounter{page}{1}
   \author{Li Zhou \inst{1,2}, Bo Ma \inst{1,2}, Yonghao Wang \inst{1,2}, Yinan Zhu \inst{3}}
   \institute{
   School of Physics and Astronomy, Sun Yat-sen University, Zhuhai 519082, China; {\it mabo8@mail.sysu.edu.cn}\\
   \and
   CSST Science Center for the Guangdong-HongKong-Macau Great Bay Area, Sun Yat-sen University, Zhuhai 519082, China;\\
   \and
   National Astronomical Observatories, Chinese Academy of Sciences, 20A Datun Road, Chaoyang District, Beijing 100012, China;
}  
   \date{Received; accepted}

\abstract{The nearby bright M-dwarf star L~98-59 has three terrestrial-sized planets. One challenge remaining in characterizing atmospheres around such planets is that it is not known a priori whether they possess any atmospheres.
Here we report the study of the atmospheres of L~98-59~c and L~98-59~d using the near-infrared spectral data from the G141 grism of HST/WFC3. 
We can reject the hypothesis of a clear atmosphere dominated by hydrogen and helium at a confidence level of $\sim$ 3~sigma for both planets. 
Thus they may have a primary hydrogen-dominated atmosphere with an opaque cloud layer, or have lost their primary hydrogen-dominated atmosphere and re-established a secondary thin atmosphere, or have no atmosphere at all. 
We cannot distinguish between these scenarios for the two planets using the current HST data. 
Future observations with JWST would be capable of confirming the existence of atmospheres around L~98-59~c and d and determining their compositions.
}
\authorrunning{Zhou et al.}
\titlerunning{Hubble WFC3 Spectroscopy of L~98-59~c \& d}
\maketitle

\section{Introduction} \label{sec:intro}
Thus far, more than 5000 exoplanets have been discovered, among which planets of 1$\sim$10 $\rm M_\oplus$ orbiting around late-type stars are the most abundant exoplanets \citep{2018AJ....156..264F, Ma18, 2021AJ....161...44E}. 
There is a dichotomy in the occurrence rate distribution of small planets: sub-Neptunes with $\rm R_p = 2.0\sim3.0$ $\rm R_\oplus$ and  super-Earths with $\rm R_p \leq 1.5$ $\rm R_\oplus$ (e.g., \citealp{2018AJ....156..264F}).
The dichotomy of occurrence rate suggests that the main composition of planets with larger radii ($\geq$ 1.8 $R_\oplus$) could be volatile (e.g., \citealp{2010ApJ...716.1208R}), whereas for smaller planets, models with more negligible atmospheres are preferred (e.g., \citealp{2015ApJ...800..135D}). 
The deficit of planets with a radius of 1.5$\sim$2.0 $\rm R_\oplus$ and short orbiting periods in the occurrence rate distribution can be explained by photo-evaporation, which may result in the stripping of the primordial $\rm H/H_e$ gas envelope \citep{2017ApJ...847...29O,2017AJ....154..109F}. There is also a deficit of massive giant planets and brown dwarfs with short orbiting periods, which is not fully understood yet \citep{Grether06, Ma14}.

One of the most important goals in exoplanets study is to search for terrestrial planets that may be habitable via detection of atmospheric bio-signatures.
Transiting exoplanet observations combined with detailed theoretical models and retrieval methods can provide valuable insights into a wide range of physical processes and chemical compositions in exoplanetary atmospheres \citep{2000ApJ...537..916S,2019ARA&A..57..617M,2022ARA&A..60..159W}.
However, due to the extremely weak signal (several tens of ppm), the atmospheric properties of rocky planets have still not been well characterized so far.

Up until now, there have only been several tentative studies of terrestrial planets atmospheres using the Hubble Space Telescope (HST)/Wide Field Camera 3 (WFC3) transmission spectrum, and most of these planets orbit nearby bright M dwarfs. 
For example, \citet{2014Natur.505...69K} and \citet{2014ApJ...794..155K} find no evident molecular absorption in the spectra of GJ~1214~b and HD~97658~b, where they rule out the cloud-free atmospheric models and suggest these planets possess cloudy atmospheres with molecules heavier than hydrogen.
The possibility of primary hydrogen-dominated clear atmospheres is also ruled out 
for seven Earth-sized planets in the TRAPPIST-1 system \citep{2016Natur.537...69D, 2018NatAs...2..214D, 2022A&A...658A.133G, 2022A&A...665A..19G}.
\cite{2016ApJ...820...99T} analyzed the transmission spectrum of a highly irradiated super-Earth, 55 Cancri e, and find that the spectrum departs from a straight line model with a 6$\sigma$ confidence level. They identified HCN as the source to explain the absorption features at 1.42 and 1.54 $\mu m$. 
\cite{2019NatAs...3.1086T} reported the first detection of water vapor in the spectrum of super-Earth K2-18~b with high statistical confidence. 
\citet{2021AJ....161...44E} find the temperate super-Earth, LHS~1140~b, could also have water vapor in its atmosphere. 
However, limited by the observation data, they cannot distinguish between the primary atmosphere model and the secondary atmosphere model \citep{2021AJ....161...44E}.  
\cite{2021AJ....161..213S} find the transmission spectrum of the highly irradiated Earth-sized planet, GJ~1132~b, shows spectral signatures of HCN and $\rm CH_4$ in an atmosphere with low mean molecular weight . 
However, \cite{2021AJ....161..284M} re-analyzed the GJ~1132~b system and did not find any molecular signatures claimed by \cite{2021AJ....161..213S}, 
which demonstrates the difficulty of using HST data to study the atmospheres of rocky planets.  
With the upcoming new facilities, such as James Webb Space Telescope  \citep[JWST;][]{2016ApJ...817...17G, 2022arXiv220811692T, 2022arXiv220708889W,  2022PASP..134i5003H, 2022MNRAS.514.2073C, 2022A&A...661A..83B} and Ariel \citep{2018ExA....46..135T}, the study of atmospheres of rocky planets will no doubt enter into a new era. 

L~98-59 is a nearby (10.6 pc) bright (K = 7.1 mag) M3 dwarf star. 
The Transiting Exoplanet Survey Satellite (TESS) has discovered three terrestrial planets in this system \citep{2019AJ....158...32K, 2019A&A...629A.111C}.
They all have a radius less than 1.6~$\rm R_\oplus$ and an orbital period ranging from 2.25 to 7.45~days. 
\citet{2021A&A...653A..41D} found evidence for a fourth non-transiting planet, and a possible fifth non-transiting small planet in the system. 
The atmospheric properties of the innermost planet L~98-59~b have been studied by \cite{2022AJ....164..203Z} and \cite{2022AJ....164..225D}. 
Both studies have rejected the hypothesis that L~98-59~b has a cloud-free hydrogen-dominated primary atmosphere. 
The HST observation results of planets c and d (proposal 15856, PI:Thomas
Barclay) will be reported in this paper.
The planets c and d will also be observed by JWST in 0.6$–$5~$\mu m$ in the Cycle 1. Thus L~98-59 is set to be one of best characterized planetary system with multiple terrestrial planets. 

In this study, we present our analysis of the atmospheric properties of the two outer most planets, L~98-59~c and L~98-59~d, using the near-infrared transmission spectrum obtained with HST/WFC3 G141 grism. 
The outline of this paper is arranged as follows. 
We first describe our data analysis in Sec~\ref{method}, which includes raw data reduction, white and spectral light curves extraction using Iraclis, and atmospheric retrievals using TauREx. 
The main results and discussion are shown in Sec~\ref{sec:results}. 
In the last section, we give a conclusion of our atmospheric study on planets L~98-59~c and L~98-59~d.

\section{Data Analysis}
\label{method}
\subsection{Data Reduction} \label{sec:analysis}
We download public available transiting observation data of L~98-59~c and L~98-59~d from Mikulski Archive for Space Telescope (MAST, proposal 15856, PI:Thomas Barclay). 
The near-infrared spatially scanned spectroscopic images were obtained using the HST/WFC3 G141 grism on April 7th, 2020 (for L~98-59~c) and January 19th and 20th, 2021 (for L~98-59~d), respectively. The images were obtained with 512 $\times$ 512 sub-array with an exposure time of 69.62s 
The scan rate is 0.496$''$ per second and the total scan length is about 37.92$''$.     
We adopt the open source pipeline Iraclis\footnote{\url{https://github.com/ucl-exoplanets/Iraclis}} to reduce the observations and extract the white and spectral light curves from raw spatially scanned spectroscopic images of HST/WFC3 \citep{2016ApJ...832..202T, 2016ApJ...820...99T}. 
The raw spectroscopic data are first processed using the following steps: zero-read subtraction, reference pixels correction, non-linearity correction, dark current subtraction, gain conversion, sky background subtraction, calibration, flat-field correction, bad pixels removal, and cosmic-ray correction. 
We refer the reader to \citep{2016ApJ...832..202T} for more details. 

\begin{figure*}[t]
\includegraphics[width=1.0\textwidth]{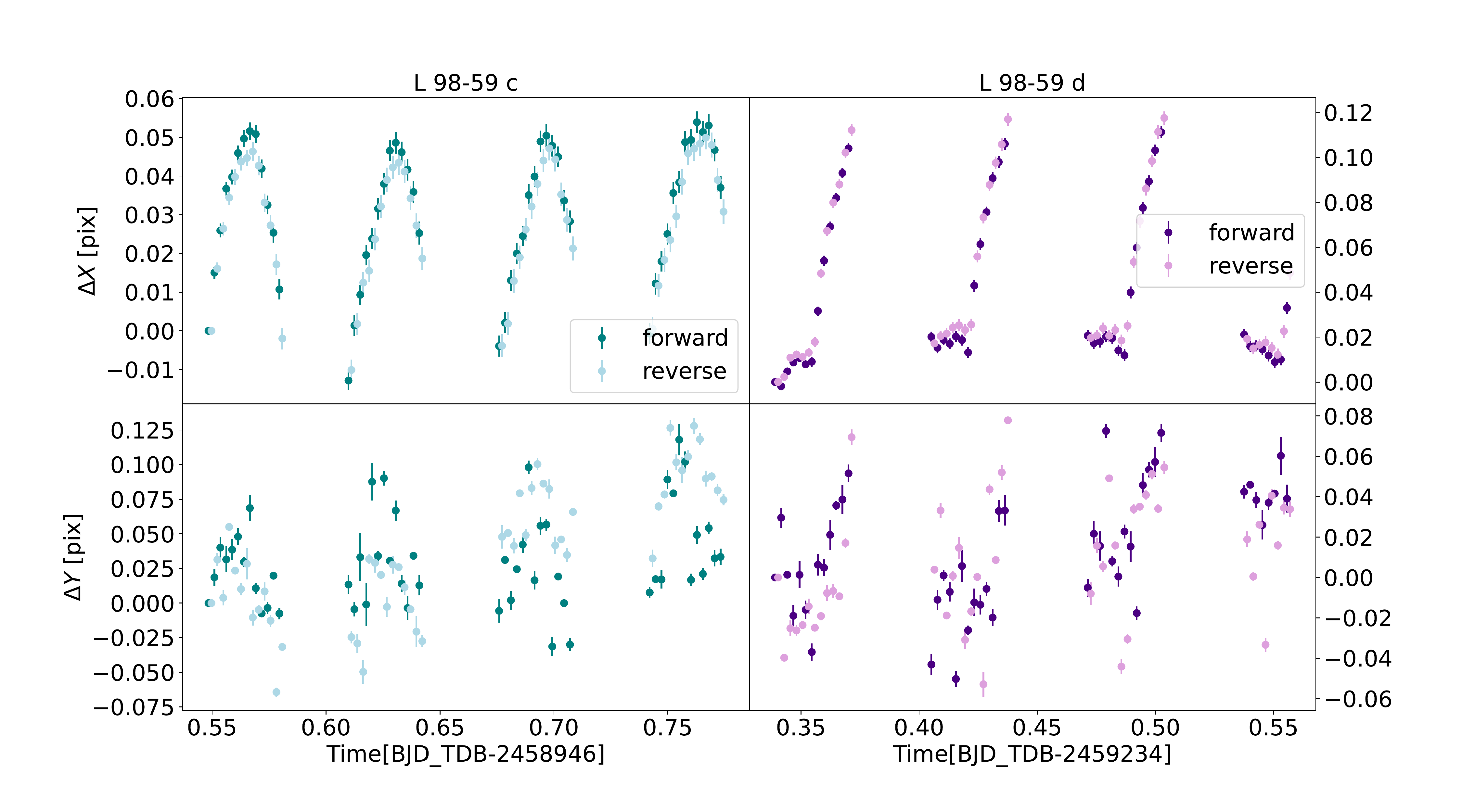}
\caption{Horizontal (top panels) and Vertical (bottom panels) shifts for each frame observation of L~98-59~c and L~98-59~d. The heavier colors represent forward scans and lighter colors represent reverse scan. All shifts are calculated against the first frame in each scan. }
\label{shift_iraclis}
\end{figure*}

\begin{figure*}[t]
\includegraphics[width=1.0\textwidth]{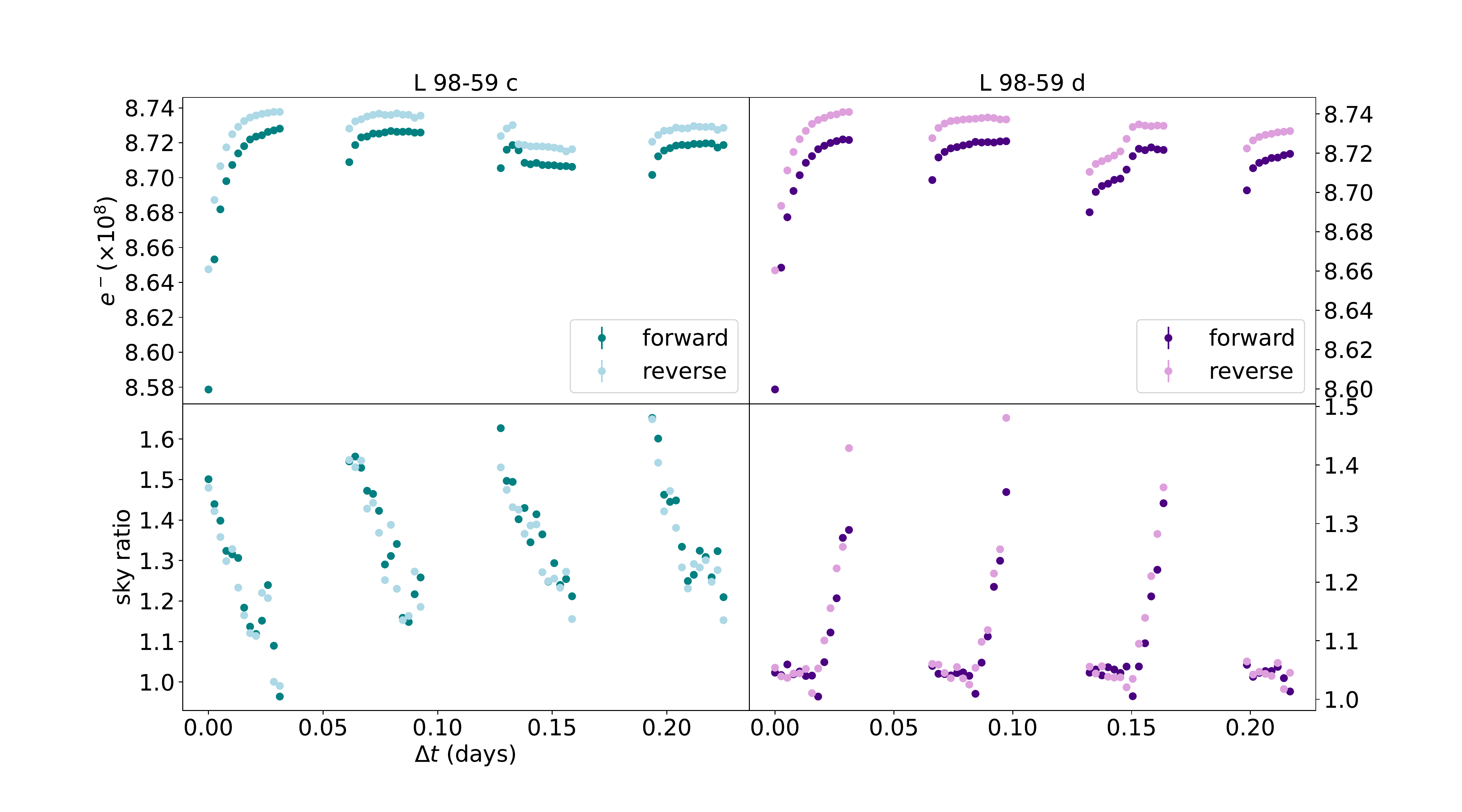}
\caption{Top panels: raw white light curves for forward scans (heavier colors) and reverse scan (lighter colors) of L~98-59~c and L~98-59~d. Bottom panels: sky background relative to the master sky frame. }
\label{flux_iraclis}
\end{figure*}

We exclude the first orbit for each observation in our data analysis to avoid the stronger wavelength-dependent ramp. When extracting white and spectral light curves from reduced scanned spectroscopic images, we take into account the geometric distortions caused by the tilted detector of the WFC3 infrared channel. 
The horizontal and vertical shifts are calculated against the first frame in each scan, and are shown in Figure~\ref{shift_iraclis}. 
We find shifts on the scale of about 0.1 pixels during the visit, which are not significant enough to affect the final transmission spectrum. 
We then obtain the white light curves by integrating across the full wavelength range of WFC3/G141 (1.088 $\sim$1.68 $\mu$m). The raw white light curves and sky ratio variation during the visit (sky background relative to the master-sky frame) are shown in Figure~\ref{flux_iraclis}. 
We also obtain a set of spectral light curves using the default `high' resolution from Iraclis as the wavelength bins, corresponding to a resolving power of 70 at 1.4 $\mu$m. 
We adopt the non-linear formula proposed in \citet{2000A&A...363.1081C}, together with the stellar models of Phoenix \citep{2018A&A...618A..20C} and stellar parameters of L~98-59 listed in Table~\ref{parameters} \citep{2021A&A...653A..41D} to calculate the limb darkening coefficients. 

\begin{figure*}[t]
\includegraphics[width=1.0\textwidth]{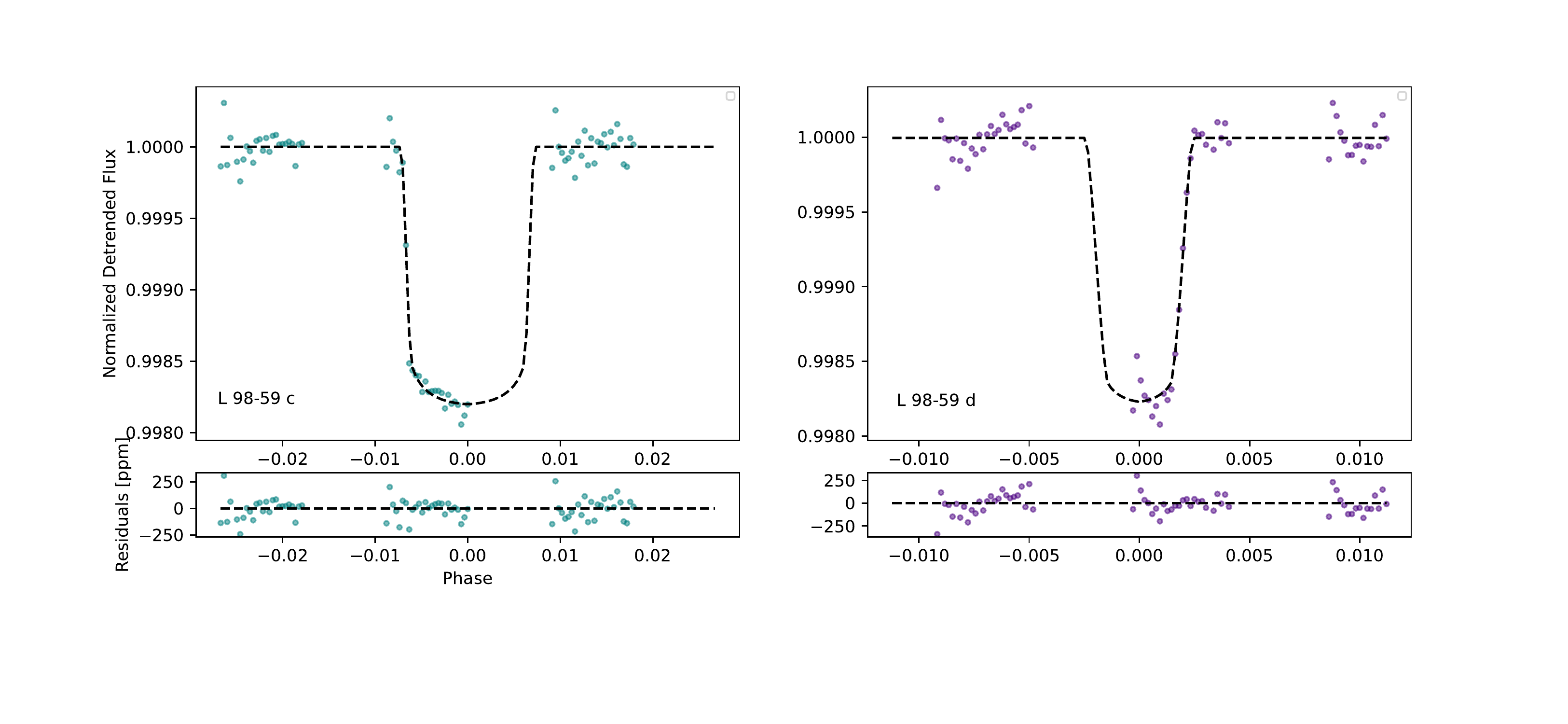}
\caption{White light curves (top panels) and model fitting residuals (bottom panels) of L~98-59~c and L~98-59~d.}
\label{lcs_iraclis}
\end{figure*}

We extract the white and spectral light curves by taking into account the time-dependent systematic noises of long-term and short-term ramps, which can be characterized by a linear trend and an exponential trend, respectively. 
The systematics are removed by multiplying a normalization factor and an instrumental correction factor.  
For the white light curves fitting, the systematics are fitted by 
\begin{equation}
 R_w(t) = n_w(1-r_a(t-T_0))(1-r_{b1}e^{-r_{b2}(t-t_0)}),
\label{white_equation}
\end{equation}
where $n_w$ is the normalization factor, which is $n_w^{for}$ for upward scanning directions and $n_w^{rev}$ for downward scanning directions. 
In this equation, t, $T_0$ and $t_0$ are time, the mid-transit time and the beginning time of each orbit, respectively. 
Also, $r_a$ is the linear trend coefficient during each visit and $r_{b1}$ and $r_{b2}$ are exponential trend coefficients along each orbit \citep{2016ApJ...832..202T,2016ApJ...820...99T}.  
When fitting the white light curves, we set the mid-transit time and $R_p/R_*$ as free parameters. 
Other parameters, such as inclination, $a/R_*$ and stellar parameters are fixed, which are listed in Table~\ref{parameters}. 
The fitted white light curves for L~98-59~c and L~98-59~d are shown in Figure~\ref{lcs_iraclis}.  

\begin{table*}[t]
\centering
\caption{Parameters of L~98-59~c and L~98-59~d applied in Iraclis and TauRex.  All these parameters were taken from \citet{2021A&A...653A..41D}.}
\label{parameters}
\begin{threeparttable}
\resizebox{0.7\textwidth}{!}{
\begin{tabular}[b]{ccc}
\hline\hline
\multicolumn{3}{c}{Stellar Parameters}\\
\multicolumn{3}{c}{L 98-59}\\
\hline
 $\rm [Fe/H](dex)$  & \multicolumn{2}{c}{-0.46 $\pm$ 0.26}\\
 $\rm T_{eff}(K)$ & \multicolumn{2}{c}{3415 $\pm$ 135}\\
 $\rm M_*(M_{\odot})$ & \multicolumn{2}{c}{0.273 $\pm$ 0.030}\\
 $\rm R_*(R_\odot)$ & \multicolumn{2}{c}{0.303 $^{+0.026}_{-0.023}$}\\
 $\rm log(g_*)(cgs)$ & \multicolumn{2}{c}{4.86 $\pm$ 0.13}\\
\hline\hline
\multicolumn{3}{c}{Planetary  Parameters}\\
\hline
&L 98-59 c & L98-59 d\\
\hline
$\rm T_{eq}(K)$ & 553 $^{+27}_{-26}$ & 416 $\pm$ 20 \\
$\rm M_p(M_\oplus)$ & 2.22 $^{+0.26}_{-0.25}$ & 1.94 $\pm$ 0.28 \\
$\rm R_p(R_\oplus)$ & 1.385 $^{+0.095}_{-0.075}$ & 1.521 $^{+0.119}_{-0.098}$\\
\hline\hline
\multicolumn{3}{c}{Transit  Parameters}\\
\hline
&L 98-59 c & L98-59 d\\
\hline
$\rm T_0(BJD)$ & 2458367.27375 $^{+0.00013}_{-0.00022}$ & 2458362.73974 $^{+0.00031}_{-0.00040}$ \\
$\rm P(days)$ & 3.6906777 $^{+0.0000016}_{-0.0000026}$ & 7.4507245 $^{+0.0000081}_{-0.0000046}$ \\
$\rm R_p/R_*$ & 0.04088 $^{+0.00068}_{-0.00056}$ & 0.04480 $^{+0.00106}_{-0.00100}$ \\
$\rm a(AU)$ & 0.0304 $^{+0.0011}_{-0.0012}$ & 0.0486 $^{+0.0018}_{-0.0019}$\\
$\rm a/R_*$ & 19.00 $^{+1.2}_{-0.8}$ & 33.7 $^{+1.9}_{-1.7}$ \\
$\rm i(deg)$ & 88.11 $^{+0.36}_{-0.16}$ & 88.449 $^{+0.058}_{-0.111}$ \\
\hline
\hline
\end{tabular}}
\end{threeparttable}
\end{table*}

\begin{figure*}[t]
\includegraphics[width=1.0\textwidth]{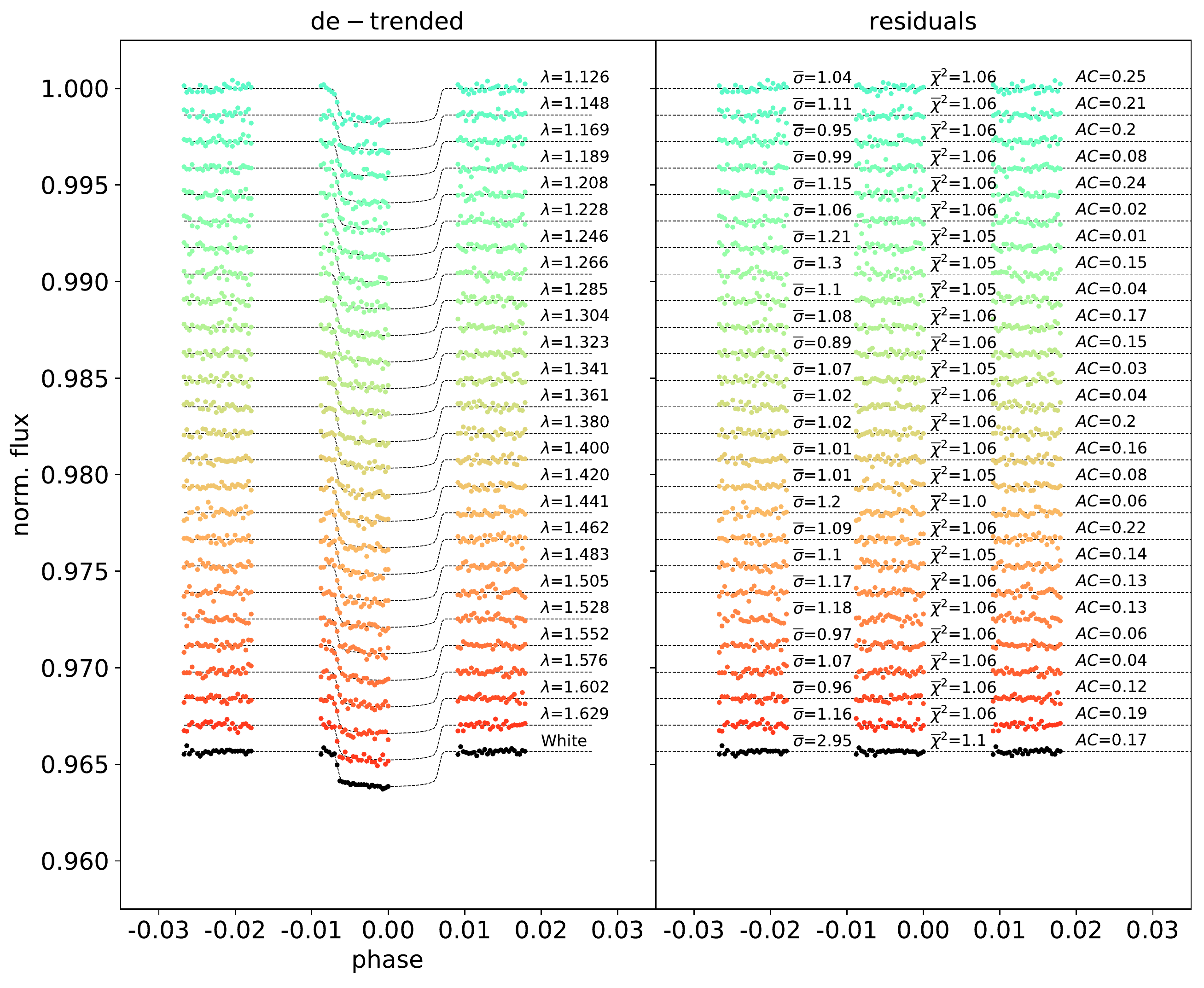}
\caption{The spectral light curve fits for the transmission spectra of L~98-59~c. For clarity, we applied an artificial offset in the y-axis. Left panel: the de-trended spectral light curves with the best fit model plotted using dotted lines. Right panels: The residuals from the fitting with mean values for the chi-squared ($\overline{\chi^2}$), the standard deviation with respect to the photon noise ($\overline{\sigma}$) and the auto-correlation (AC).}
\label{all_fitting_c}
\end{figure*}

\begin{figure*}[t]
\includegraphics[width=1.0\textwidth]{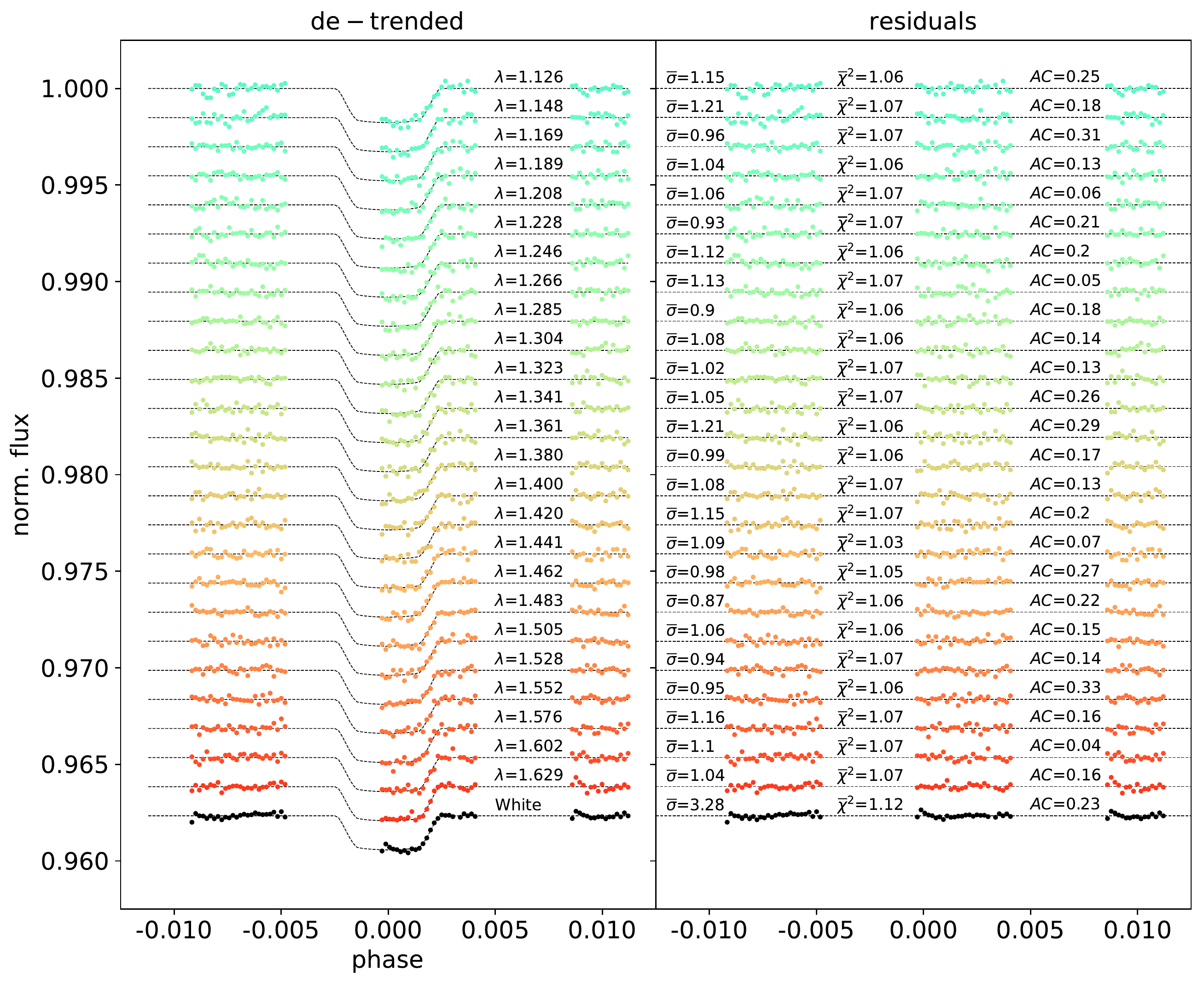}
\caption{Same as Fig~\ref{all_fitting_c}, but for L~98-59~d.}
\label{all_fitting_d}
\end{figure*}

For spectral light curves, we fit the systematics by 
\begin{equation}
 R_\lambda(t) = n_\lambda(1-r_\lambda(t-T_0))(LC_w/M_w),
\end{equation}
where $r_\lambda$ is the wavelength-dependent linear systematic trend coefficient along each visit, $LC_w$ is the white light curve and $M_w$ is the best-fit white light curve model. 
Similar with Equation~\ref{white_equation},  $n_\lambda$ is $n_\lambda^{for}$ for upward scanning directions and $n_\lambda^{rev}$ for downward scanning directions. 
In the process of spectral light curve fitting, we set $R_p/R_*$ as the only free parameter. Other planetary parameters, stellar parameters and orbit parameters are fixed. 
The fitting spectral light curves for L~98-59~c and L~98-59~d are shown in Figure~\ref{all_fitting_c} and Figure~\ref{all_fitting_d}, respectively. 
The left panels present de-trended spectral light curves together with the best-fit models at each wavelength bin. 
Right panels present the standard deviation of the residuals with respect to the photon noise ($\overline{\sigma}$), the reduced chi-square of residuals from the fitting ($\overline\chi^2$), and the auto-correlation function ($\rm AC$) during the transits. 
The final transmission spectra are obtained and shown in Figure~\ref{spectrum_cd}, with a mean uncertainty of about 45 and 57 ppm for L~98-59~c and L~98-59~d, respectively. From the figure, we cannot see strong absorption features in their transmission spectra for both L~98-59~c and L~98-59~d. 

\subsection{Atmospheric Retrieval} \label{sec:retrieve}
We use TauREx3\footnote{\url{https://github.com/ucl-exoplanets/TauREx3_public}} \citep{2015ApJ...802..107W, 2021ApJ...917...37A} to perform atmospheric retrieval analysis for L~98-59~c and L~98-59~d. 
TauREx3 adopts the nested sampling algorithm Multinest \citep{2008MNRAS.384..449F, 2009MNRAS.398.1601F,2011ascl.soft09006F, 2014A&A...564A.125B, 2019OJAp....2E..10F} to map the atmospheric parameter space and find the best fit model for the observed transmission spectrum with a Bayesian analysis framework. 

In our retrieval analysis, we set the planetary radius in the range of 0.6 $\rm R_p$ to 1.6 $\rm R_p$ as the priors, the equilibrium temperature of the planet in the range of 0.6 $\rm T_{eq}$ to 1.6 $\rm T_{eq}$, the tolerance evidence to 0.5, and the number of live points to 1500 for the nested sampling. 
$\rm R_p$ (1.385 $R_\oplus$ for L~98-59~c and 1.521 $R_\oplus$ for L~98-59~d) and $\rm T_{eq}$ (553K for L~98-59~c and 416 K for L~98-59~d) are the prior values taken from Table~\ref{parameters}.   
Considering the narrow wavelength range of our observed data, we perform the atmospheric retrieval using an isothermal temperature-pressure profile with 100 plane-parallel layers, with pressure varying from $10 ^{-3}$ to $10^6$ Pa. The molecular abundance is assumed to be constant in each layer, which does not vary with altitude. 

The absorption trace gases included in the model are taken from ExoMol \citep{2016JMoSp.327...73T,2021A&A...646A..21C}, HITEMP \citep{2010JQSRT.111.2139R} and HITRAN \citep{1987ApOpt..26.4058R}, such as $\rm H_2O$ \citep{2018MNRAS.480.2597P},
$\rm CO_2$ \citep{2010JQSRT.111.2139R},
$\rm CO$ \citep{2015ApJS..216...15L}, 
$\rm CH_4$ \citep{2014MNRAS.440.1649Y}, $\rm NH_3$ \citep{2019MNRAS.490.4638C}, 
$\rm HCN$ \citep{2014MNRAS.437.1828B}, $\rm TiO$ \citep{2019MNRAS.488.2836M}, $\rm VO$ \citep{2016MNRAS.463..771M}. We also include collision-induced absorption (CIA) of $\rm H_2$-$\rm H_2$ \citep{doi:10.1021/jp109441f,2018ApJS..235...24F}, $\rm H_2$-$\rm He$ \citep{doi:10.1063/1.3676405} and Rayleigh scattering. 
The grey clouds \citep{2013ApJ...778...97L} are included with the top pressure of clouds ranging from $10 ^{-3}$ to $10^6$ Pa. 
We introduce $\rm H_2$, $\rm He$, and $\rm N_2$ as filling gases since they only contribute to the mean molecular weight and do not display absorption features in the spectrum. 
We fix the ratio between $\rm He$ and $\rm H_2$ abundance to the solar value of 0.17. 

We perform both of primary and secondary atmospheric model retrievals. 
In the hydrogen-rich primary atmospheric model, we allow the ratio between $\rm N_2$ and $\rm H_2$ abundance to vary from $10^{-12}$ to $10^{-2}$, and the volume mixing ratios (VMRs) of all trace gases to vary from $10^{-12}$ to $10^{-1}$. 
In the secondary atmospheric model, we allow the ratio between the abundance of $\rm N_2$ and $\rm H_2$ to vary from $10^{-12}$ to $10^{4}$, and the VMRs of other gases to vary from $10^{-12}$ to 1, which can be used to model a thin atmosphere with high mean molecular weight.
We also choose to perform between cloudy and clear atmospheric retrievals by including or excluding grey clouds in the atmospheric models. 

We also conduct a flat-line atmospheric retrieval, which only includes grey clouds in the model. 
The flat-line model is used as a baseline for comparison to evaluate the significance of other models.

\section{Results and Discussion} \label{sec:results}
In this section, we will present our atmospheric modeling results, investigate possible spectral contamination from stellar activities, and discuss implication of study this system by future space missions.

\subsection{Retrieval Results}
\begin{figure*}[t]
\includegraphics[width=0.5\textwidth]{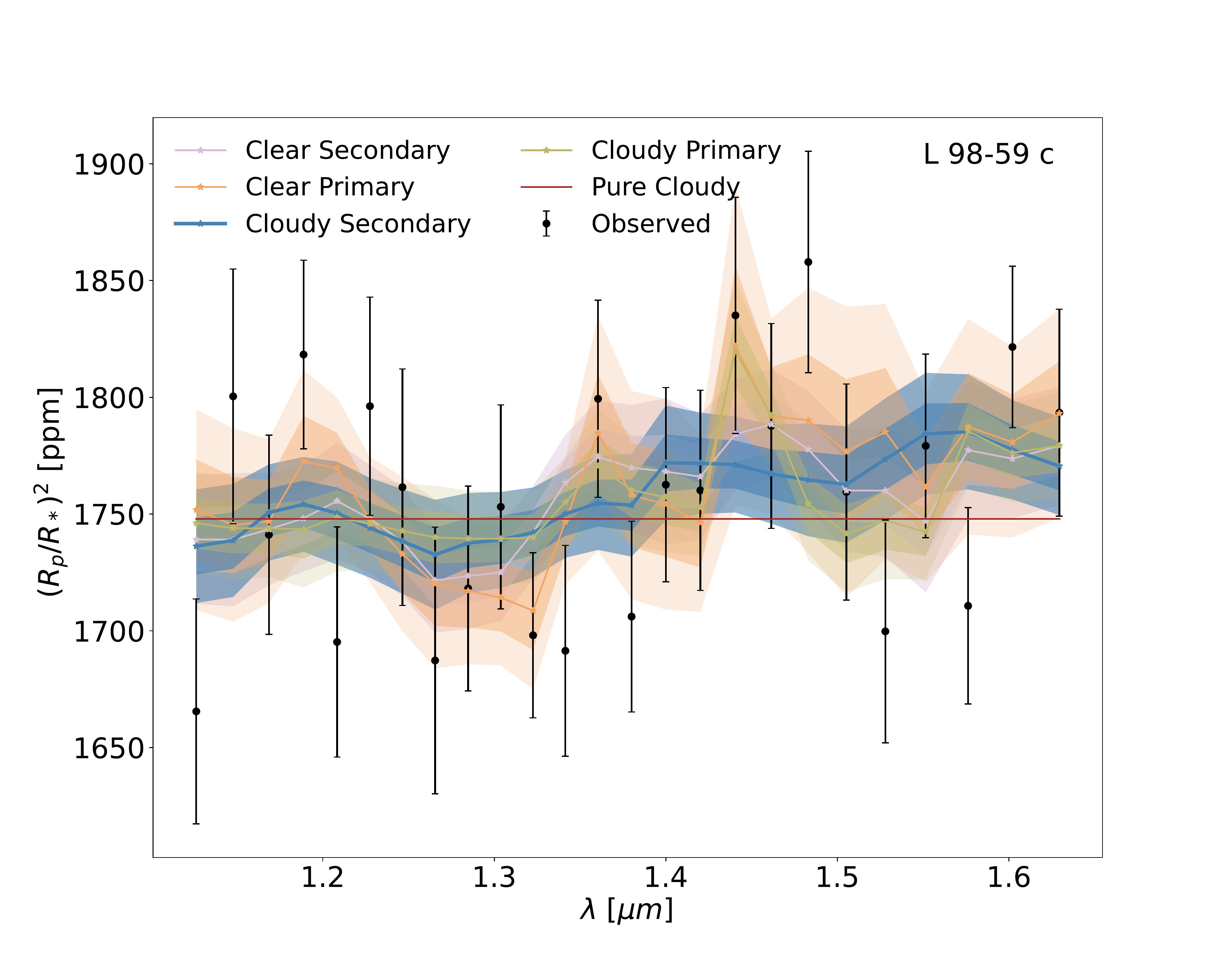}
\includegraphics[width=0.5\textwidth]{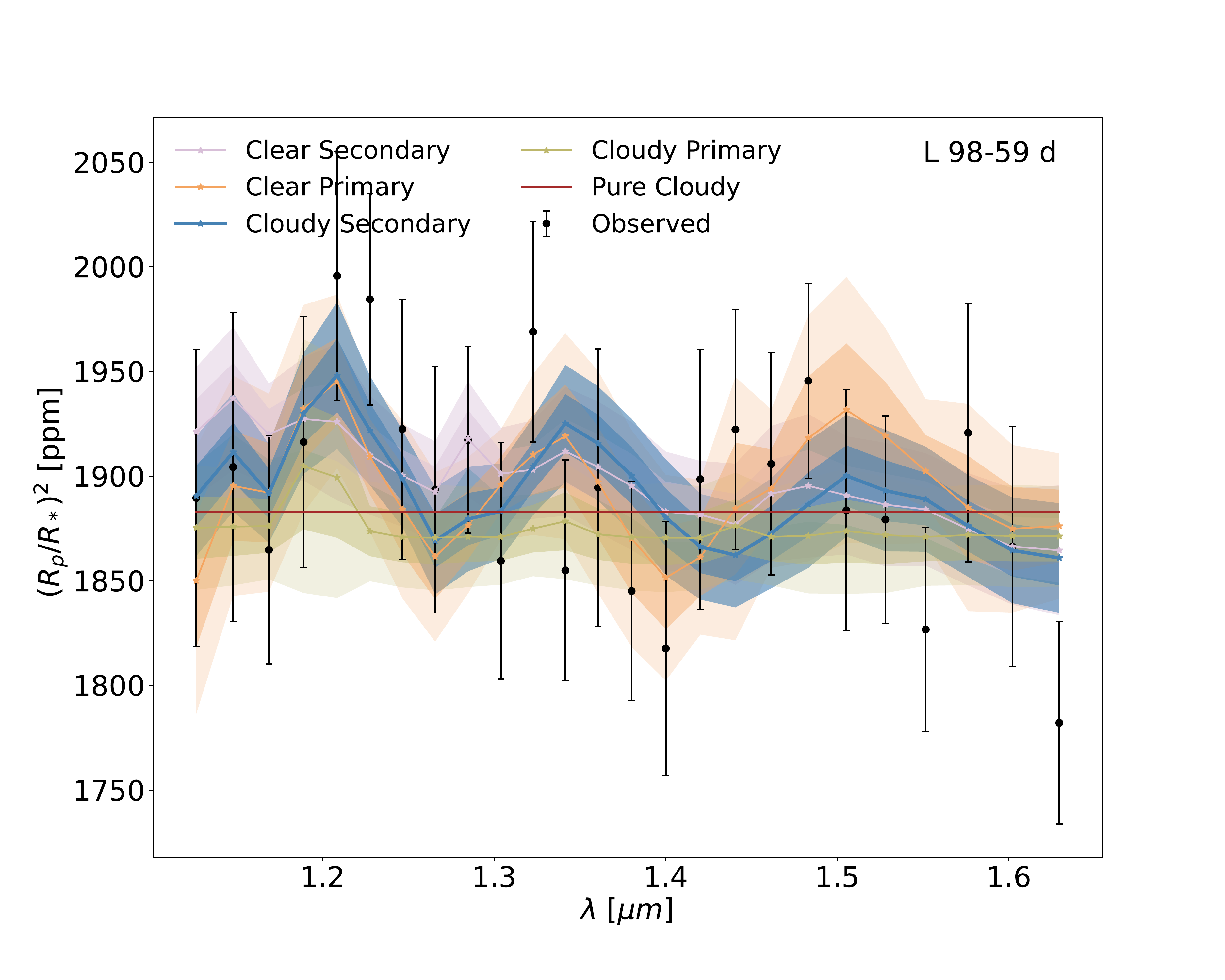}
\caption{Different atmospheric retrieval models for L~98-59~c and L~98-59~d. }
\label{spectrum_cd}
\end{figure*}

We have conducted five different model retrieval for both of L~98-59~c and L~98-59~d. 
The HST transmission spectra of L~98-59~c and L~98-59~d, together with the different model fitting results, are shown in Figure~\ref{spectrum_cd}. The 1$\sigma$ (shadow regions with heavier colors) and 2$\sigma$ (shadow regions with lighter colors) uncertainties are also over-plotted. 
We choose the flat-line pure cloudy model as the baseline for model comparison and use the Bayes factor $\Delta log_{10}(E)$ \citep{Kass1995bayes} and n-$\sigma$ values \citep{2013ApJ...778..153B} to evaluate the goodness of model fitting, where $\rm \Delta log_{10}(E) = log_{10}(E_{models})- log_{10}(E_{flat~model})$ is the logarithmic ratio of Bayesian evidences between models with trace gas molecules and the flat-line model. 
The Bayesian evidence refers to the fully marginalized likelihood
\citep{Marshall06, Trotta07, Nelson20}, and the Bayes factor can be interpreted against empirical scales \citep{Kass1995bayes}. 
According to Table~2 of \citet{2013ApJ...778..153B}, a Bayes factor value greater than 2.5 means a moderate detection, and a factor value greater than 5.0 indicates a strong detection. 
Similarly, a n-$\sigma$ value between 2.1$\sigma$ and 2.7$\sigma$ indicates a weak detection, a n-$\sigma$ value greater than 2.7$\sigma$ indicates a moderate detection, and a n-$\sigma$ value greater than 3.6$\sigma$ indicates a strong detection \citep{2013ApJ...778..153B}.
The statistical results and model parameters, including $\Delta log_{10}(E)$, n-$\sigma$ value, the planet radius, equilibrium temperature, cloud top pressure, mean molecular weight, and scale heights, are shown in Table~\ref{retrieval_results} and Table~\ref{retrieval_pars}. 

From the statistical values shown in Table~\ref{retrieval_results}, we can reject the clear primary model for both planets. For the other four models, they all show similar Bayesian evidence values and we cannot be sure which one of them represents the true nature of these two planets.
Therefore, both of L~98-59~c and L~98-59~d could be lack of an atmosphere, or the atmosphere could be a thin secondary atmosphere with high molecular weight, or a primary atmosphere with a high photochemical haze layer. 

All the model posterior distributions are shown in the Appendix (Figure~\ref{c_clear_primary} to Figure~\ref{d_cloudy}). We generally can find an anti-correlation in the planet radius-equilibrium temperature panel and a positive correlation in the planet radius-cloud top pressure panel in the corner plots.
Next we will present all the retrieval results for these five atmospheric models.

\begin{table*}[t]
\centering
\caption{Statistical values of different models: Bayesian evidences and Sigma. }
\label{retrieval_results}
\begin{threeparttable}
\resizebox{\textwidth}{!}{
\begin{tabular}[b]{lllllll}
\hline
model & descriptions & VMR priors & $\rm log_{10}$(E) & $\rm \Delta log_{10}$(E) & Sigma\\
\hline
&&&&&\\
&&L~98-59~c&&&\\
&&&&&\\
\hline
clear primary & CO/$\rm CO_2$/$\rm NH_3$/$\rm CH_4/$ & $10^{-12}$ $\sim$ $10^{-1}$ for trace gases & 202.543 $\pm$ 0.076 & -3.017 & -2.956 $\sigma$\\
& $\rm H_2O$/HCN/$\rm T_iO$/VO+$\rm N_2$/$\rm H_2$ & $10^{-12}$ $\sim$ $10^{-2}$ for $\rm N_2$/$\rm H_2$ & & & \\
\hline
cloudy primary & clouds+CO/$\rm CO_2$/$\rm NH_3$/$\rm CH_4$/ & $10^{-12}$ $\sim$ $10^{-1}$ for trace gases & 205.047 $\pm$ 0.060 & -0.513 & -1.682 $\sigma$\\
& $\rm H_2O$/HCN/$\rm T_iO$/VO+$\rm N_2$/$\rm H_2$ & $10^{-12}$ $\sim$ $10^{-2}$ for $\rm N_2$/$\rm H_2$ & & & \\
\hline
clear secondary & CO/$\rm CO_2$/$\rm NH_3$/$\rm CH_4/$ & $10^{-12}$ $\sim$ 1 for other gases & 205.843 $\pm$ 0.059 & 0.283 & 1.468 $\sigma$\\
& $\rm H_2O$/HCN/$\rm T_iO$/VO+$\rm N_2$/$\rm H_2$ & $10^{-12}$ $\sim$ $10^4$ for $\rm N_2$/$\rm H_2$ & & & \\
\hline
cloudy secondary & clouds+CO/$\rm CO_2$/$\rm NH_3$/$\rm CH_4$/ & $10^{-12}$ $\sim$ 1 for other gases & 205.841 $\pm$ 0.057 &  0.281 & 1.466 $\sigma$\\
& $\rm H_2O$/HCN/$\rm T_iO$/$\rm VO$+$\rm N_2$/$\rm H_2$ & $10^{-12}$ $\sim$ $10^{4}$ for $\rm N_2$/$\rm H_2$ & & & \\
\hline
Pure cloudy & clouds & - & 205.560 $\pm$ 0.057 & - & -\\
\hline
\hline
&&&&&\\
&&L~98-59~d&&&\\
&&&&&\\
\hline
clear primary & CO/$\rm CO_2$/$\rm NH_3$/$\rm CH_4/$ & $10^{-12}$ $\sim$ $10^{-1}$ for trace gases & 202.870 $\pm$ 0.068 & -2.642 & -2.811 $\sigma$\\
& $\rm H_2O$/HCN/$\rm T_iO$/VO+$\rm N_2$/$\rm H_2$ & $10^{-12}$ $\sim$ $10^{-2}$ for $\rm N_2$/$\rm H_2$ & & & \\
\hline
cloudy primary & clouds+CO/$\rm CO_2$/$\rm NH_3$/$\rm CH_4$/ & $10^{-12}$ $\sim$ $10^{-1}$ for trace gases & 205.131 $\pm$ 0.059 & -0.381 & -1.566$\sigma$\\
& $\rm H_2O$/HCN/$\rm T_iO$/VO+$\rm N_2$/$\rm H_2$ & $10^{-12}$ $\sim$ $10^{-2}$ for $\rm N_2$/$\rm H_2$ & & & \\
\hline
clear secondary & CO/$\rm CO_2$/$\rm NH_3$/$\rm CH_4/$ & $10^{-12}$ $\sim$ 1 for other gases & 205.794 $\pm$ 0.059 &  0.282 &  1.467 $\sigma$\\
& $\rm H_2O$/HCN/$\rm T_iO$/VO+$\rm N_2$/$\rm H_2$ & $10^{-12}$ $\sim$ $10^4$ for $\rm N_2$/$\rm H_2$ & & & \\
\hline
cloudy secondary & clouds+CO/$\rm CO_2$/$\rm NH_3$/$\rm CH_4$/ & $10^{-12}$ $\sim$ 1 for other gases & 205.898 $\pm$ 0.056 & 0.386 &  1.571 $\sigma$\\
& $\rm H_2O$/HCN/$\rm T_iO$/$\rm VO$+$\rm N_2$/$\rm H_2$ & $10^{-12}$ $\sim$ $10^{4}$ for $\rm N_2$/$\rm H_2$ & & & \\
\hline
Pure cloudy & clouds & - & 205.512 $\pm$ 0.056 &  - & -\\
\hline
\hline
\end{tabular}}
\end{threeparttable}
\end{table*}

\begin{table*}[t]
\centering
\caption{The retrieval parameters of different models for L~98-59~c and L~98-59~d.}
\label{retrieval_pars}
\begin{threeparttable}
\resizebox{\textwidth}{!}{
\begin{tabular}[b]{lllcllll}
\hline
model &$\rm R_p (R_{Jup})$ & T (K) & $\rm log_{10}(P) (Pa)$ & $\mu$ ($\rm g~mol^{-1}$) & H (km)  & H (ppm)\\
\hline
&&&&&&\\
&&&L~98-59~c&&&\\
&&&&&&\\
\hline
clear primary & 0.12$^{+0.00}_{-0.00}$ & 473.57$^{+304.81}_{-103.65}$ & - & 2.32$^{+2.28}_{-0.01}$ & 141.14 & 54.70 \\
\hline
cloudy primary & 0.10$^{+0.01}_{-0.01}$ & 512.85$^{+178.95}_{-114.50}$ & 1.03$^{+1.90}_{-1.92}$ & 2.32$^{+0.16}_{-0.02}$ & 106.15 & 41.13\\
\hline
clear secondary & 0.12$^{+0.00}_{-0.00}$ & 620.82$^{+172.11}_{-179.25}$ & - & 27.42$^{+1.31}_{-6.14}$ & 15.66 & 6.07 \\
\hline
cloudy secondary &  0.12$^{+0.00}_{-0.00}$ & 609.02$^{+163.17}_{-173.07}$ & 2.68$^{+2.12}_{-2.94}$ & 27.12$^{+1.17}_{-8.86}$ & 15.53 & 6.02\\
\hline
Pure cloudy & 0.11$^{+0.01}_{-0.01}$ & 556.55$^{+210.25}_{-168.26}$ & 2.20$^{+2.64}_{-3.28}$ & 2.30$^{+0.00}_{-0.00}$ & 140.59 & 54.48 \\
\hline
\hline
&&&&&&\\
&&&L~98-59~d&&&\\
&&&&&&\\
\hline
clear primary & 0.12$^{+0.00}_{-0.00}$ & 296.19$^{+58.63}_{-33.15}$ & - & 2.49$^{+4.01}_{-0.18}$ & 94.12 & 39.97 \\
\hline
cloudy primary & 0.11$^{+0.01}_{-0.01}$ & 400.2$^{+143.21}_{-94.39}$ & 1.17$^{+1.69}_{-2.03}$ & 2.33$^{+0.27}_{-0.02}$ & 114.20 & 48.50 \\
\hline
clear secondary & 0.13$^{+0.00}_{-0.00}$ & 461.85$^{+121.67}_{-125.02}$ & - & 27.81$^{+4.42}_{-7.22}$ & 15.42 & 6.55\\
\hline
cloudy secondary & 0.13$^{+0.00}_{-0.00}$ & 449.96$^{+128.86}_{-119.79}$ & 2.42$^{+2.23}_{-2.91}$ & 27.33$^{+2.87}_{-10.40}$ & 15.29 & 6.49 \\
\hline
Pure cloudy & 0.11$^{+0.01}_{-0.01}$ & 433.32$^{+159.13}_{-130.98}$ & 2.25$^{+2.83}_{-3.43}$ & 2.30$^{+0.00}_{-0.00}$ & 125.26 & 53.20 \\
\hline
\hline
\end{tabular}}
\end{threeparttable}
\end{table*}

\paragraph{\noindent \textbf{Clear Primary}}
The posterior distribution of clear primary atmosphere model parameters for L~98-59~c and L~98-59~d are presented in Figure~\ref{c_clear_primary} and Figure~\ref{d_clear_primary}. From Table~\ref{retrieval_results}, the statistical significance of the clear primary atmosphere models is 2.96$\sigma$ and 2.81$\sigma$lower than the flat-line models for L~98-59~c and L~98-59~d respectively.
Therefore, we can reject the clear primary atmosphere models for both planets.

\paragraph{\noindent \textbf{Cloudy Primary}}
The posterior distributions of cloudy primary atmosphere 
models for L~98-59~c and L~98-59~d are presented in Figure~\ref{c_cloudy_primary} and Figure~\ref{d_cloudy_primary}. 
According to the posterior distributions, the planet radii at 10~bar are $\rm 0.10^{+0.01}_{-0.01}\; R_J$ and $\rm 0.11^{+0.01}_{-0.01}\; R_J$, the equilibrium temperatures are $\rm 512.85^{+178.95}_{-114.50}\;K$ and $\rm 400.20^{+143.21}_{-94.39} \; K$, respectively.
In the primary atmospheric models, the ratio of $\rm N_2/H_2$ is allowed to vary between $10^{-12}$ and $10^{-2}$. 
The best-fitted value of $\rm log(N_2/H_2)$ is $\rm -6.67^{+2.80}_{-3.12}$ and $\rm -7.26^{+2.93}_{-2.73}$ for L~98-59~c and L~98-59~d, respectively. 
We find the peak of $\rm log(P_{clouds})$ distribution locates at about $\rm 1.03^{+1.90}_{-1.92}$ for L~98-59~c and $\rm 1.17^{+1.69}_{-2.03}$ for L~98-59~d, which corresponds to a top cloud layer pressure at about $\rm 10^{-4}$~bar for these two planets. 
The clouds at this pressure layer may be photochemical mists or hazes with big particles, but not condensation clouds \citep{2019ARA&A..57..617M, 2022A&A...658A.133G}.

\paragraph{\noindent \textbf{Clear Secondary} }
According to Table~\ref{retrieval_results}, the Bayesian evidence of cloud-free secondary atmosphere model is also similar to that of the flat-line model for both planets.
We present their posterior distributions in Figure~\ref{c_clear_secondary} and Figure~\ref{d_clear_secondary}. 
From the corner plots, we can see that for L~98-59~c and L~98-59~d, the planet radii at 10~bar are $\rm 0.12^{+0.00}_{-0.00}\; R_J$ and $\rm 0.13^{+0.00}_{-0.00}\; R_J$, the equilibrium temperatures are $\rm 620.82^{+172.11}_{-179.25}\; K$ and $\rm 461.85^{+121.67}_{-125.02}\; K$, respectively.
With the abundance of $\rm N_2$ retrieved as the mixing ratio of inactive gases, the value of $\rm log(N_2/H_2)$ is allowed to increase beyond 1 and the best-fitted value is $\rm 1.70^{+1.41}_{-1.51}$ and $\rm 1.56^{+1.54}_{-1.71}$ for L~98-59~c and L~98-59~d, respectively. 
The mean molecular weight reaches $\rm 27.42^{+1.31}_{-6.14}\; g/mol$ for L~98-59~c and $\rm 27.81^{+4.42}_{-7.22}\;g/mol$ for L~98-59~d, which is decided mostly by the $\rm N_2$ abundance. 
We can find a clear positive correlation between the mean molecular weight and $\rm N_2$ abundance from Figure~\ref{c_clear_secondary} and Figure~\ref{d_clear_secondary}. 
The corresponding scale height values are $\sim$ 16~km ($\sim$ 6.1~ppm) and $\sim$ 15~km ($\sim$ 6.6~ppm) for L~98-59~c and for L~98-59~d, respectively. 
The small scale height values make it very hard to probe the molecular abundance of these two planets using the current HST spectra, and we should treat the derived abundance values with caution.

\paragraph{\noindent \textbf{Cloudy Secondary}}
From the logarithmic Bayesian evidence ($\rm log_{10}(E)$) of the atmospheric retrieval results in Table~\ref{retrieval_results}, we can see that cloudy secondary atmosphere model is the most favored model for both L~98-59~c and L~98-59~d. 
This best-fitted model is also shown in Figure~\ref{spectrum_cd}, and the posterior distributions of the planet radius, the equilibrium temperature, the molecular volume mixing ratios, cloud top pressure, and mean molecular weight are shown in Figure~\ref{c_cloudy_secondary} and Figure~\ref{d_cloudy_secondary}.

For L~98-59~c and L~98-59~d, the atmospheric retrieval analysis for cloudy secondary atmosphere reveals the radii at 10~bar to be $\rm 0.12^{+0.00}_{-0.00} \;R_J$ and $\rm 0.13^{+0.00}_{-0.00} \;R_J$, the equilibrium temperature to be $\rm 609.02^{+163.17}_{-173.07}\;K$ and $\rm 449.96^{+128.86}_{-119.79} \;K$, respectively.
The best-fitted value of $\rm log(N_2/H_2)$ is $\rm 1.60^{+1.48}_{-1.59}$ for L~98-59~c and $\rm 1.40^{+1.60}_{-1.91}$ for L~98-59~d. 
The mean molecular weight reaches $\rm 27.12^{+1.17}_{-8.86}\; g/mol$ for L~98-59~c and $\rm 27.33^{+2.87}_{-10.40}\;g/mol$ for L~98-59~d. 
The corresponding scale height values are 16 km ($\sim$ 6.2 ppm) and 15 km ($\sim$ 6.5 ppm) for L~98-59~c and for L~98-59~d, respectively. 
The cloud top pressure parameter $\rm log(P_{clouds})$ is found to be $\rm 2.68^{+2.12}_{-2.94}$ for L~98-59~c and $\rm 2.42^{+2.23}_{-2.91}$ for L~98-59~d, which corresponds to a cloud top layer at about $\rm 10^{-2.5}$~bar for these two planets. 
At the layer with this pressure, the clouds are most likely to be condensation clouds \citep{2019ARA&A..57..617M, 2022A&A...658A.133G}. 

\paragraph{\noindent \textbf{Pure Cloudy}}
The flat-line pure cloudy model is used as the baseline for model comparison. 
A flat exoplanet transmission spectrum can generally be attributed to two different causes: a high-altitude aerosol layer blocking absorption or a high mean molecular weight atmosphere with small scale height. 
Our statistical results indicate it is also a model that can reasonably explain our current observed transmission spectra data. 

\subsection{Stellar Contamination}
For a M-type star, the stellar activity can cause spectral contamination \citep{Huerta08, Ma12}. Its impact on the transmission spectra of exoplanets need to be investigated \citep{2018ApJ...853..122R}. Here we study the potential contamination caused by stellar activity on the transmission spectra of L~98-59~c and L~98-59~d. 
L~98-59 is likely to be a quiet M-dwarf with weak XUV activity \citep{2021AJ....162..169P}, since there is no stellar activity features seen from the TESS data \citep{2019AJ....158...32K}. 
We adopt the stellar contamination models from \cite{2018ApJ...853..122R}, which include four cases: giant spots, solar-like spots, giant spots with faculae and solar-like spots with faculae. 
The angular radii are defined as $R_{giant}$ = $7^\circ$ and $R_{solar}$ = $2^\circ$, respectively.
We estimate the contamination factor, which is the effect of stellar activity on the planetary transmission spectrum, using Equation 3 from \cite{2018ApJ...853..122R}. 
The spots and faculae covering fractions for M3 type stars are used in the equation. 
The stellar flux of photosphere, spots and faculae are calculated using the theoretical BT-Settle models with temperatures of 3200, 2800 and 3300~K, respectively. 
The surface gravity of L~98-59 is set to $\rm log(g) = 5.0$ and the stellar metallicity to $\rm [Fe/H] = -0.5$.

\begin{figure*}[t]
\includegraphics[width=0.5\textwidth]{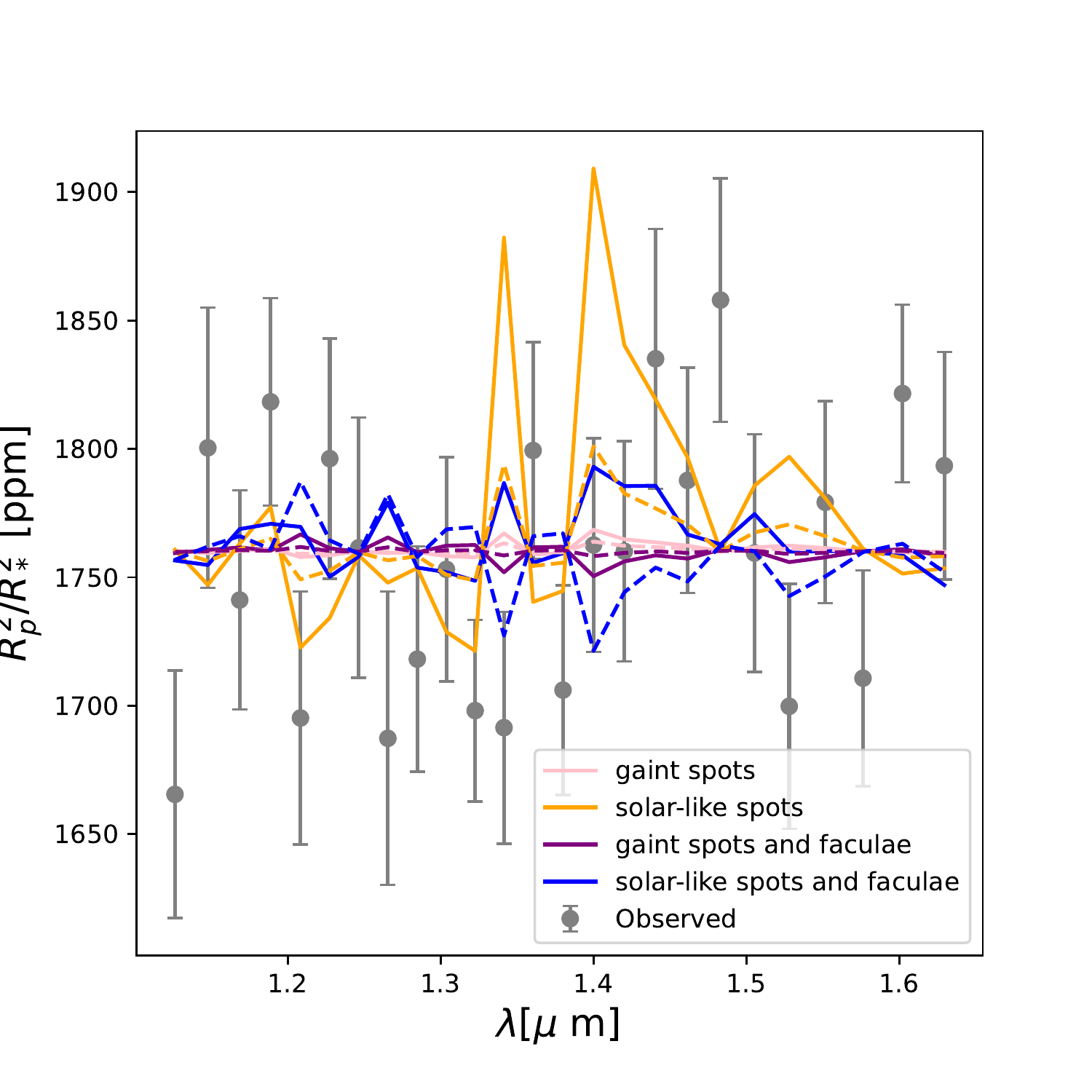}
\includegraphics[width=0.5\textwidth]{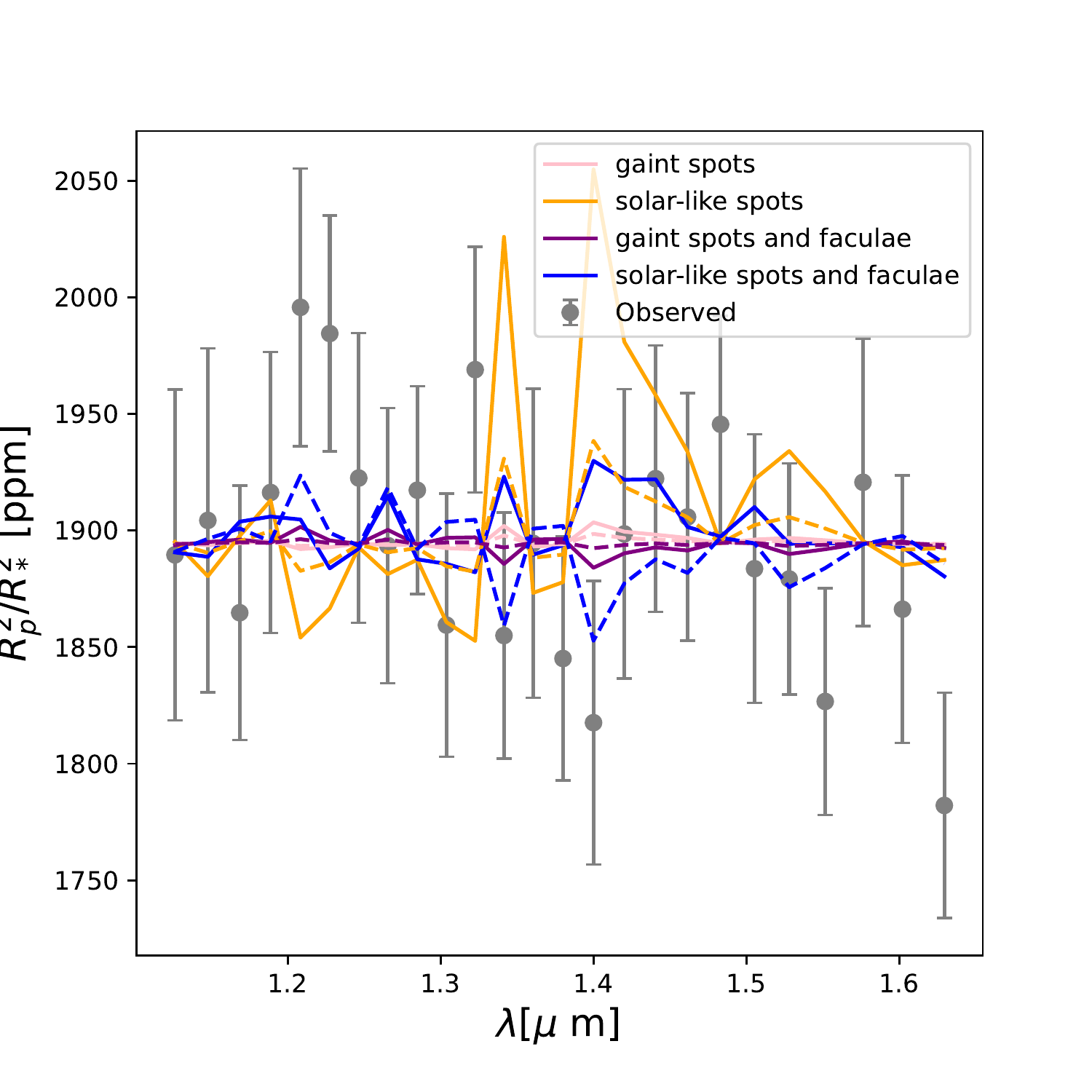}
\caption{The stellar contamination models for L~98-59~c and L~98-59~d. }
\label{stellar contamination}
\end{figure*}

By multiplying the contamination factor with a flat-line transit depth model, we derive the contamination spectra of L~98-59~c and L~98-59~d and present them in Figure~\ref{stellar contamination}. 
The lines with different colors in the figure represent different stellar contamination models.
Solid lines represent the transit light source effect for the maximum spot or faculae filling factor, and dashed lines represent mean filling factor.
We compare the stellar contamination models with the observed transmission spectra to check which model can match the observed spectra. 
We calculate the $\chi^2$ and reduced-$\chi^2$ for different model comparison in Table~\ref{chi of stellar contaminations}, which show stellar models of solar-like spots with maximum filling factor deviates most from the observed data. 

\begin{table}[t]
\centering
\caption{The $\chi^2$ and reduced-$\chi^2$ for different stellar contamination models.}
\label{chi of stellar contaminations}
\begin{threeparttable}
\resizebox{0.5\textwidth}{!}{
\begin{tabular}{l|l|l|c}
\hline
\multicolumn{2}{c|}{Stellar model} & $\chi^2$ & reduced-$\chi^2$ \\
\hline
\multicolumn{4}{c}{L~98-59~c} \\
\hline
\multirow{2}{*}{Giant spots} & max & 33.387 & 1.518 \\
\cline{2-4}
& mean & 33.356 & 1.516 \\
\hline
\multirow{2}{*}{Solar-like spots} & max & 62.618 & 2.846 \\
\cline{2-4}
& mean & 35.330 & 1.606 \\
\hline
\multirow{2}{*}{\shortstack{Giant spots\\+ faculae}} & max & 33.717 & 1.606 \\
\cline{2-4}
& mean & 33.423 & 1.592 \\
\hline
\multirow{2}{*}{\shortstack{Solar-like spots\\+ faculae}} & max & 35.680 & 1.699 \\
\cline{2-4}
& mean & 36.699 & 1.748 \\
\hline
\multicolumn{4}{c}{L~98-59~d} \\
\hline
\multirow{2}{*}{Giant spots} & max & 22.761 & 1.035 \\
\cline{2-4}
& mean & 22.225 & 1.010 \\
\hline
\multirow{2}{*}{Solar-like spots} & max & 57.154 & 2.598 \\
\cline{2-4}
& mean & 27.633 & 1.256 \\
\hline
\multirow{2}{*}{\shortstack{Giant spots\\+ faculae}} & max & 20.429 & 0.973 \\
\cline{2-4}
& mean & 21.475 & 1.023 \\
\hline
\multirow{2}{*}{\shortstack{Solar-like spots\\+ faculae}} & max & 24.579 & 1.170 \\
\cline{2-4}
& mean & 17.555 & 0.836 \\
\hline
\hline
\end{tabular}}
\end{threeparttable}
\end{table}

To study the impact of stellar contamination on atmospheric retrieval results, we decide to subtract the potential contamination of stellar activity from the planetary transmission spectra and make additional atmospheric retrievals using the resultant spectra assuming different atmospheric models: cloudy primary, cloudy secondary, clear primary, clear secondary, and flat-line model. 
We present the retrieval results in Table~\ref{stellar contaminations lgE}. 
We can see that for all the models (except for solar-like spots with maximum filling factor), the statistical results are similar to those before the stellar contamination removal. 
The clear primary atmosphere model can be excluded with a $\sim$ 3$\sigma$ significance, which still suggests these two planets may have no atmosphere, or have a thin secondary generation of atmosphere, or a primary hydrogen dominated atmosphere with an opaque cloud layer.

\begin{table*}
\centering
\footnotesize
\caption{The retrieval results in the case of different stellar contamination models.}
\label{stellar contaminations lgE}
\begin{tabular}{lllllll}
\hline
\multicolumn{2}{c}{\multirow{3}*{Stellar model}} & \multirow{1}*{log(E) } & \multirow{1}{*}{log(E) } & \multirow{1}{*}{log(E)  } & \multirow{1}{*}{log(E) } & \multirow{1}{*}{log(E) } \\
\multicolumn{2}{c}{} & ($\Delta$ log(E), Sigma) & ($\Delta$ log(E), Sigma) & ($\Delta$ log(E), Sigma) & ($\Delta$ log(E), Sigma) & \\ 
\multicolumn{2}{c}{} & cloudy primary & cloudy secondary & clear primary & clear secondary & flat \\
\hline
&&&&&&\\
\multicolumn{7}{c}{L~98-59~c}\\
&&&&&&\\
\hline
\multirow{2}{*}{Giant spots} & max & 205.333 (-0.403, -1.59$\sigma$) & 206.145 (0.409, 1.59$\sigma$) & 202.733 (-3.003, -2.95$\sigma$) & 205.994 (0.258, 1.44$\sigma$) & 205.736
\\
\cline{3-7}
& mean & 205.149 (-0.557, -1.72$\sigma$) & 205.994 (0.288, 1.47$\sigma$) & 202.632 (-3.074, -2.98$\sigma$) & 205.779 (0.073, 1.18$\sigma$) & 205.706  \\
\hline
\multirow{2}{*}{Solar-like spots} & max & 201.178 (-0.371, -1.56$\sigma$) & 201.748 (0.199, 1.37$\sigma$) & 200.715 (-0.834, -1.92$\sigma$) & 201.537 (-0.012, -1.01$\sigma$) & 201.549 \\
\cline{3-7}
& mean & 205.516 (-0.579, -1.74$\sigma$) & 206.274 (0.179, 1.35$\sigma$) & 203.134 (-2.961, -2.93$\sigma$) & 206.056 (-0.039, -1.10$\sigma$) & 206.095 \\
\hline
\multirow{2}{*}{\shortstack{Giant spots\\+faculae}} & max & 204.757 (-0.386, -1.57$\sigma$) & 205.369 (0.226, 1.40$\sigma$) & 202.168 (-2.975, -2.94$\sigma$) & 205.483 (0.34, 1.53$\sigma$) & 205.143 \\
\cline{3-7}
& mean & 205.076 (-0.42, -1.60$\sigma$) & 205.606 (0.11, 1.24$\sigma$) & 202.405 (-3.091, -2.98$\sigma$) & 205.814 (0.318, 1.50$\sigma$) & 205.496 \\
\hline
\multirow{2}{*}{\shortstack{Solar-like spots \\+faculae}} & max & 205.412 (-0.664, -1.80$\sigma$) & 206.207 (0.131, 1.28$\sigma$) & 202.980 (-3.096, -2.98$\sigma$) & 206.202 (0.126, 1.27$\sigma$) & 206.076  \\
\cline{3-7}
& mean & 202.256 (-0.477, -1.65$\sigma$) & 203.324 (0.591, 1.74$\sigma$) & 200.440 (-2.293, -2.67$\sigma$) & 203.594 (0.861, 1.94$\sigma$) & 202.733  \\
\hline
None & - & 205.114 (-0.398, -1.58$\sigma$) & 205.801 (0.289, 1.48$\sigma$) & 202.293 (-3.219, -3.03$\sigma$) & 205.772 (0.26, 1.44$\sigma$) & 205.512 \\
\hline\hline
&&&&&&\\
\multicolumn{7}{c}{L~98-59~d}\\
&&&&&&\\
\hline
\multirow{2}{*}{Giant spots} & max & 204.609 (-0.713, -1.83$\sigma$)  & 205.364 (0.042, 1.11$\sigma$) & 202.64 (-2.682, -2.83$\sigma$) & 205.479 (0.157, 1.32$\sigma$) & 205.322
\\
\cline{3-7}
& mean & 204.967 (-0.476, -1.65$\sigma$) & 205.754 (0.311, 1.50$\sigma$) & 202.686 (-2.757, -2.86$\sigma$) & 205.578 (0.135, 1.28$\sigma$) & 205.443  \\
\hline
\multirow{2}{*}{Solar-like spots} & max & 197.207 (0.139, 1.29$\sigma$) & 197.405 (0.337, 1.52$\sigma$) & 197.560 (0.492, 1.66$\sigma$) & 197.645 (0.577, 1.73$\sigma$) & 197.068  \\
\cline{3-7}
& mean & 203.442 (-0.4, -1.58$\sigma$) & 204.010 (0.168, 1.33$\sigma$) & 201.847 (-1.995, -2.54$\sigma$) & 204.056 (0.214, 1.39$\sigma$) & 203.842 \\
\hline
\multirow{2}{*}{\shortstack{Giant spots\\+faculae}} & max & 205.488 (-0.62, -1.77$\sigma$) & 206.457 (0.349, 1.54$\sigma$) & 202.980 (-3.128, -3.0$\sigma$) & 206.251 (0.143, 1.30$\sigma$) & 206.108  \\
\cline{3-7}
& mean & 205.120 (-0.584, -1.74$\sigma$) & 205.967 (0.263, 1.45$\sigma$) & 202.797 (-2.907, -2.91$\sigma$) & 205.885 (0.181, 1.35$\sigma$) & 205.704  \\
\hline
\multirow{2}{*}{\shortstack{Solar-like spots \\+faculae}} & max & 204.735 (-0.481, -1.66$\sigma$) & 205.523 (0.307, 1.49$\sigma$) & 203.220 (-1.996, -2.54$\sigma$) & 205.355 (0.139, 1.29$\sigma$) & 205.216  \\
\cline{3-7}
& mean & 206.476 (-0.811, -1.90$\sigma$) & 207.489 (0.202, 1.37$\sigma$) & 203.592 (-3.695, -3.20$\sigma$) & 207.256 (-0.031, -1.08$\sigma$) & 207.287  \\
\hline
None & - & 205.043 (-0.488, -1.66$\sigma$) & 205.851 (0.32, 1.51$\sigma$) & 202.684 (-2.847, -2.89$\sigma$) & 205.655 (0.124, 1.27$\sigma$) & 205.531 \\
\hline\hline
\end{tabular}
\end{table*}


\subsection{Future Missions}
In this work, each planet has only been observed at one transiting window. Thus more observations from HST/WFC3 G141 can provide more reliable and precise spectral results.
In addition, observations from JWST would also be very powerful in the study of  the planetary atmosphere models \citep{2022arXiv220708889W, 2022arXiv220512972H, 2022MNRAS.514.2073C, 2022AJ....164..225D}. 
JWST can provide unprecedented observations of exoplanet atmospheres due to its unmatched infrared sensitivity, higher spectral resolution and wider wavelength coverage, which will enable detection of more subtle molecular absorption features in the atmosphere. 

\begin{figure*}[t]
\includegraphics[width=0.5\textwidth]{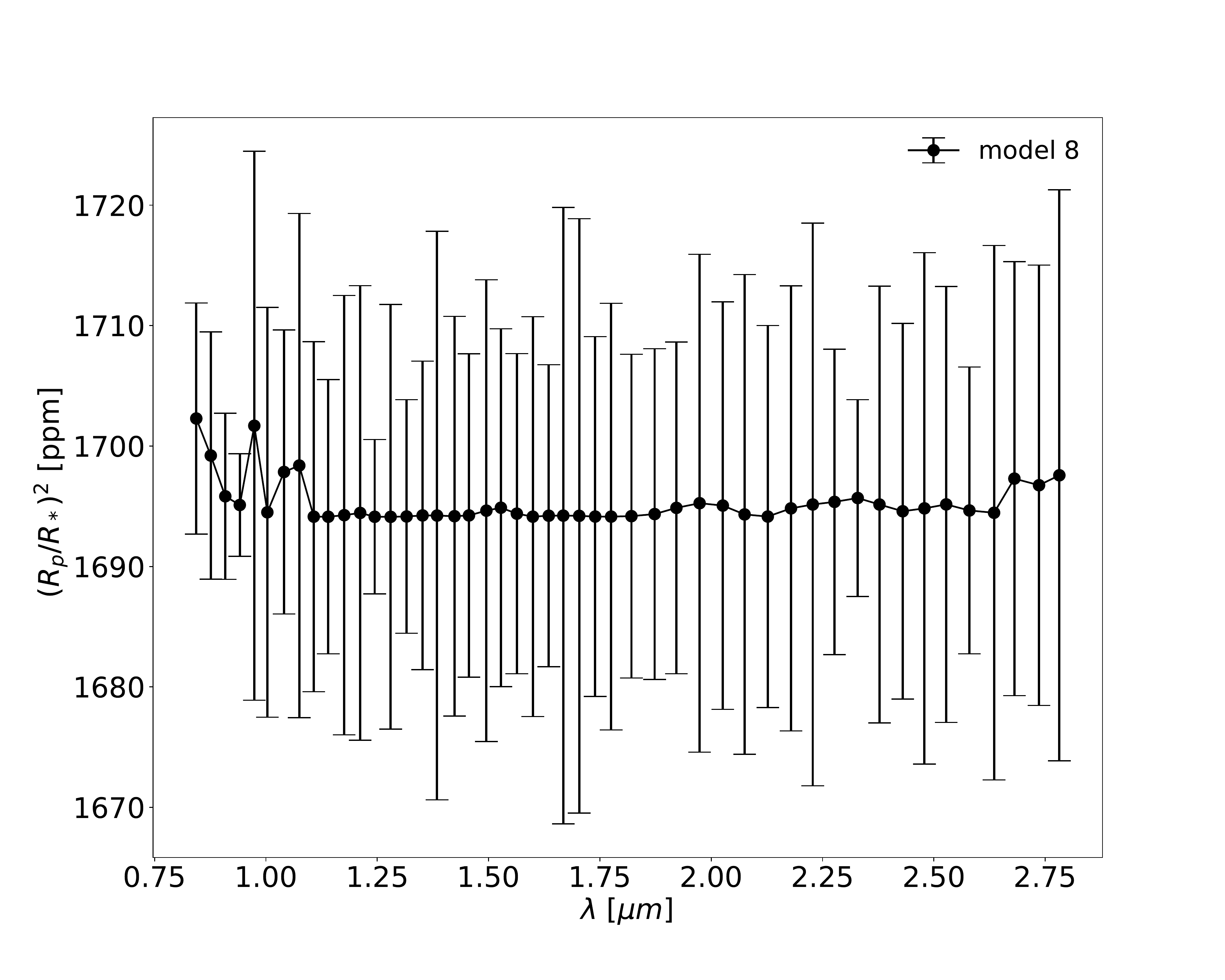}
\includegraphics[width=0.5\textwidth]{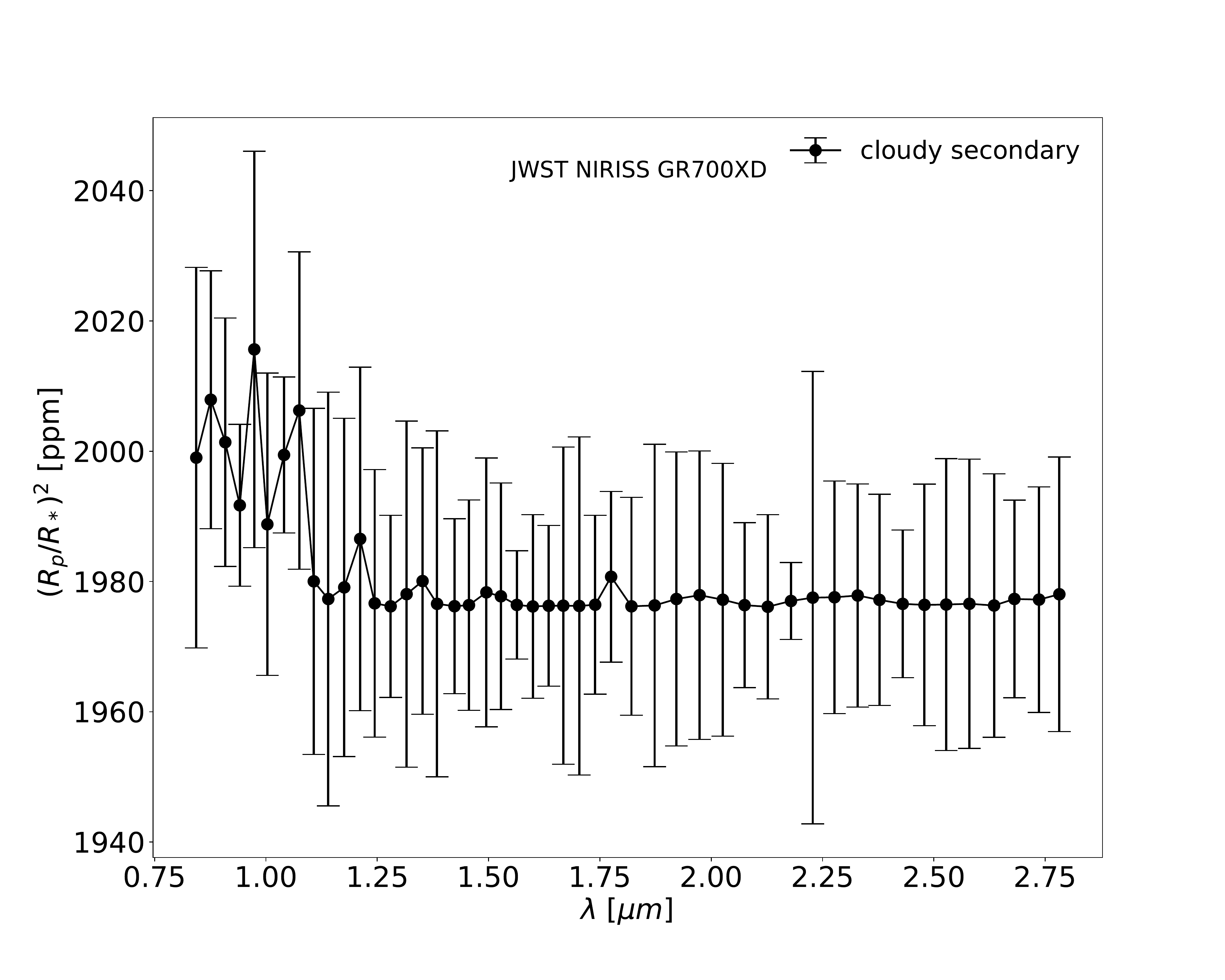}
\caption{Theoretic spectral model of L~98-59~c (left panel) and L~98-59~d (right panel) in the 
wavelength range of JWST/NIRISS GR700XD, together
with the predicted noise in one single transit visit run. }
\label{jwst}
\end{figure*}

JWST will observe planets c and d in the Cycle 1.
Here we generate the theoretical transmission spectra of L~98-59~c and L~98-59~d from one single transit using the NIRISS GR700XD spectrograph on JWST, assuming a cloudy secondary atmospheric model for both planets. 
The simulated spectra are shown in Figure~\ref{jwst}. We can clearly see that there are no strong molecular absorption features in the wavelength range of JWST/NIRISS, which agrees with a flat-line spectral model.
By comparing the future observations from JWST with these simulated spectra, we will be able to put more constraints on the planetary atmospheric models. Thus, we strongly recommend JWST to cover these small rocky planets in its observation plan. 

\section{Conclusions} \label{sec:conclusion}
Characterizing the atmospheric properties of terrestrial planets is a very important topic in exoplanet study nowadays.
In this paper, we present the near-infrared transmission spectra obtained with HST/WFC3 G141 grism for the terrestrial planets L~98-59~c (1.4~$\rm R_\oplus$) and L~98-59~d (1.5~$\rm R_\oplus$), which are the two outer planets in the multi-planet system L~98-59. 
We have also studied the possible impacts of contamination from stellar activity.
We find similar atmospheric retrieval results for both planets. 
We can reject the model of a clear primordial atmosphere dominated by hydrogen with about 3$\sigma$ confidence for both planets. 
From the similar Bayesian evidences obtained when fitting the other four atmospheric models, we conclude that these two planets may either have no atmosphere, or have a secondary generation thin atmosphere with a high molecular weight, or have a primary atmosphere with a thick opaque cloud layer. 
Future measurements from JWST can be used to put more constraints on their atmospheric properties, and demonstrate which scenario is more appropriate.

Knowing the diversity of planetary processes is important for the understanding of our planet Earth. Thus \cite{2021AJ....162..169P} advocate that this transiting multi-planet system is an excellent target for comparative planetary study. 
By investigating the transmission spectra of three transiting planets which are formed in the same protoplanetary disk and orbiting the same star, the system-wide trends in the atmospheric composition and volatile retention can be revealed \citep{2022AJ....164..225D}. 
\cite{2021AJ....162..169P} have evaluated the detectability of certain spectral features from the three transiting terrestrial planets using simulated transmission spectral observation from HST and JWST. 
Based on the empirically determined cosmic shoreline, the relative size and insolation flux of the L~98-59~b planet places it within the regime of significant atmospheric loss \citep{Zahnle17, 2021AJ....162..169P}. 
Thus L~98-59~b is likely to have lost its primary atmosphere caused by efficient hydrodynamic escape \citep{2021A&A...653A..41D}. 
Theoretical calculations suggest that the retention of a pure-$\rm H_2O$ atmosphere is also difficult because of a high escape efficiency \citep{Johnstone20}. All these scenarios agree well with the HST observations \citep{2022AJ....164..203Z, 2022AJ....164..225D}.
L~98-59~c and d are also near the cosmic shoreline, and have likely experienced a run-away greenhouse phase \citep{2021AJ....162..169P, 2021A&A...653A..41D}. Their higher masses may have inhibited mass loss, leaving them likely with a Venus-like atmosphere dominated by $\rm CO_2$. 
However, unlike JWST \citep{2022arXiv220811692T, 2022arXiv221110489A}, HST does not have the power to detect $CO_2$, which has a strong opacity near 4.3$\mu m$. 
Our analysis of the HST data in this paper has ruled out the possibility of a hydrogen dominated clear atmosphere for both planets c and d. 
A secondary atmosphere from volcanic outgassing or volatile retention such as $\rm CO_2$ cannot be completely ruled out \citep{Kite20,  2021AJ....161..213S, 2022AJ....164..225D}, which is consistent with this post-runaway greenhouse scenario. 
Future confirmation of volatile absorption features, such as $\rm CO_2$, in these planets can help to understand the planet-mass-dependent atmospheric retention of terrestrial planets \citep{2021A&A...653A..41D, 2021AJ....162..169P, 2022AJ....164..225D}.

\begin{acknowledgements}
We thank professor Giovana Tinetti, whose visit has triggered this study. 
We thank Angelos Tsiaras, Ingo Waldmann and Ahmed Al-Refaie for their instructions on how to use Iraclis and Taurex. 
We acknowledge the financial support from the National Key R\&D Program of China (2020YFC2201400), NSFC grant 12073092, 12103097, 12103098, 11733006, the science research grants from the China Manned Space Project(No. CMS-CSST-2021-B09), Guangzhou Basic and Applied Basic Research Program (202102080371), China Postdoctoral Science Foundation (No.2020M672936), and the Fundamental Research Funds for the Central Universities, Sun Yat-sen University.

This work is based on observations with the NASA/ESA Hubble Space Telescope, obtained at the Space Telescope Science Institute (STScI) operated by AURA, Inc. The publicly available HST observations presented here were taken as part of proposal 15856, led by Thomas Barclay. These were obtained from the Hubble Archive, which is part of the Mikulski Archive for Space Telescopes. 

\end{acknowledgements}

 {   \nocite{*}
     \bibliography{l98-59cd.bib}
     \bibliographystyle{aasjournal}
    }

\appendix
\section{Additional figures}
\begin{figure*}[t]
\includegraphics[width=1.0\textwidth]{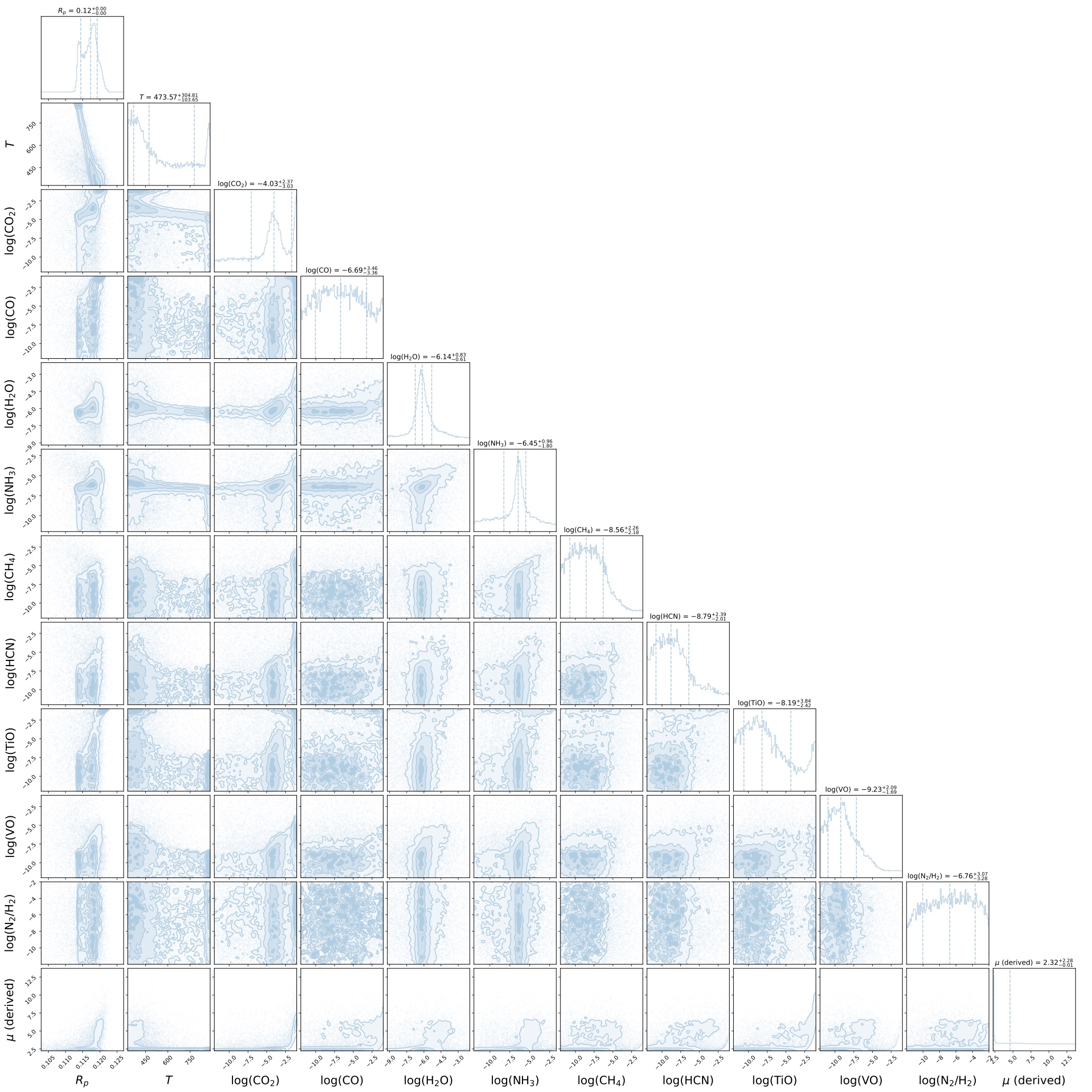}
\caption{The atmospheric retrieval posterior distributions of the clear primary atmosphere model for L 98-59 c.}
\label{c_clear_primary}
\end{figure*}
\begin{figure*}[t]
\includegraphics[width=1.0\textwidth]{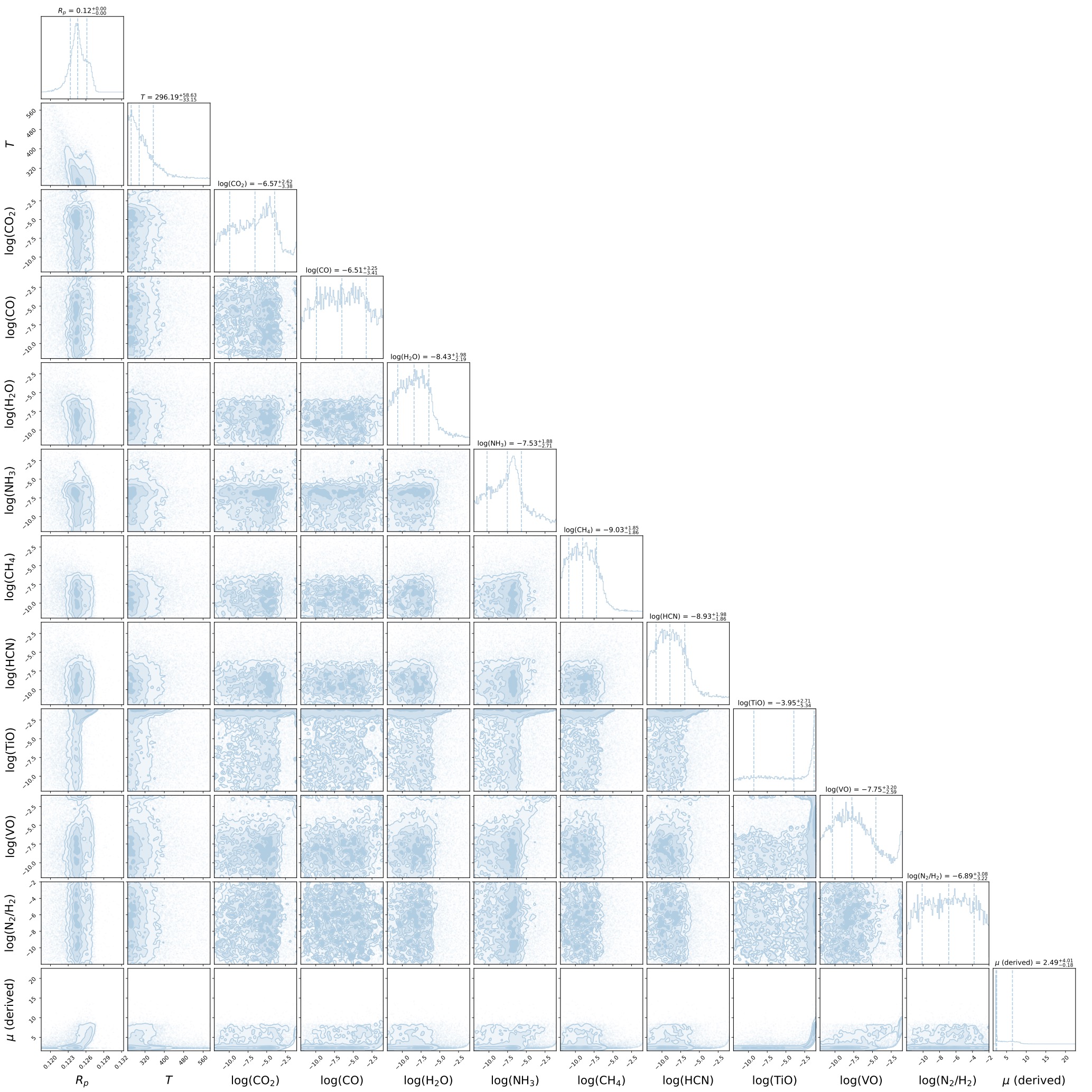}
\caption{The atmospheric retrieval posterior distributions of the clear primary atmosphere model for L~98-59~d.}
\label{d_clear_primary}
\end{figure*}
\begin{figure*}[t]
\includegraphics[width=1.0\textwidth]{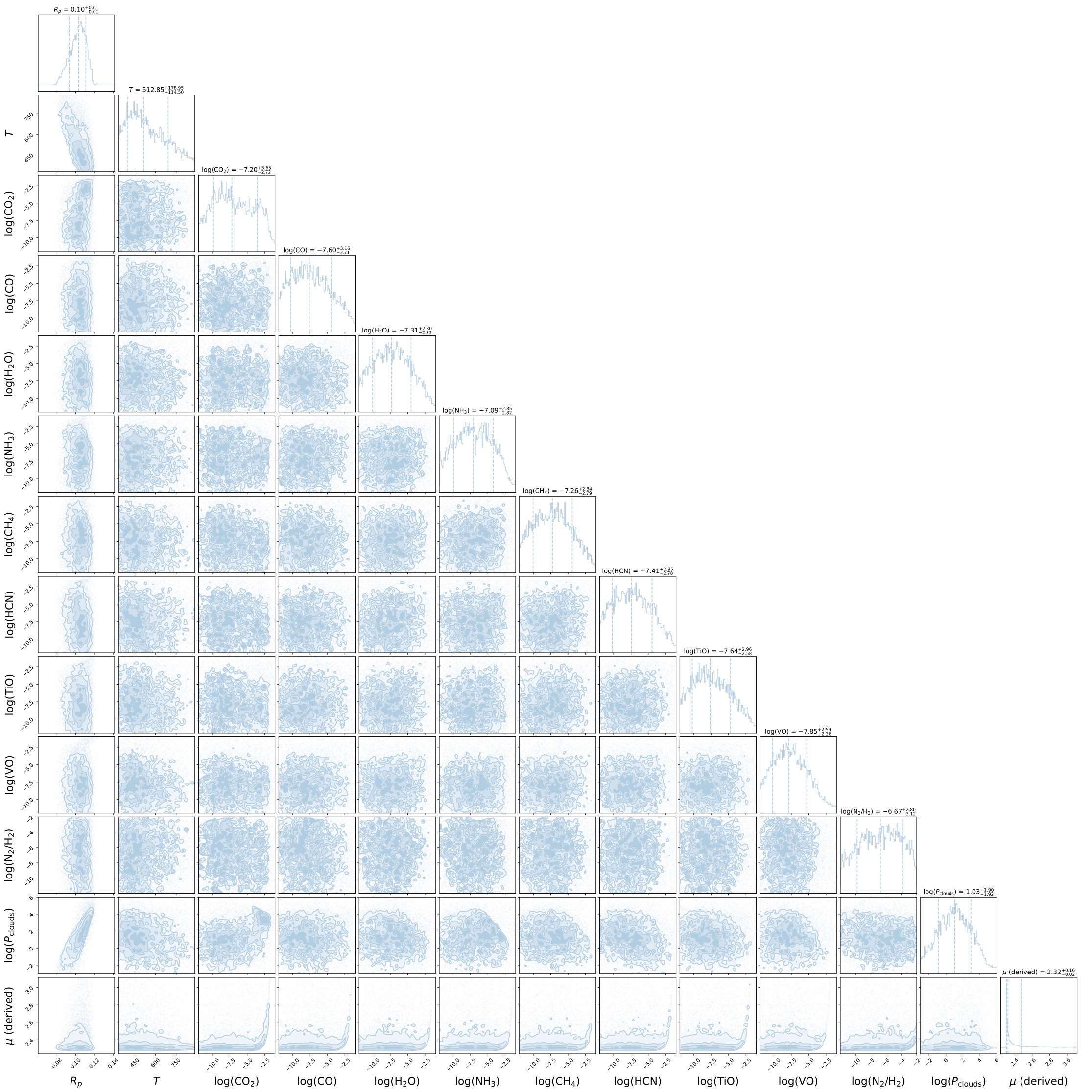}
\caption{The atmospheric retrieval posterior distributions of the cloudy primary atmosphere model for L 98-59 c.}
\label{c_cloudy_primary}
\end{figure*}
\begin{figure*}[t]
\includegraphics[width=1.0\textwidth]{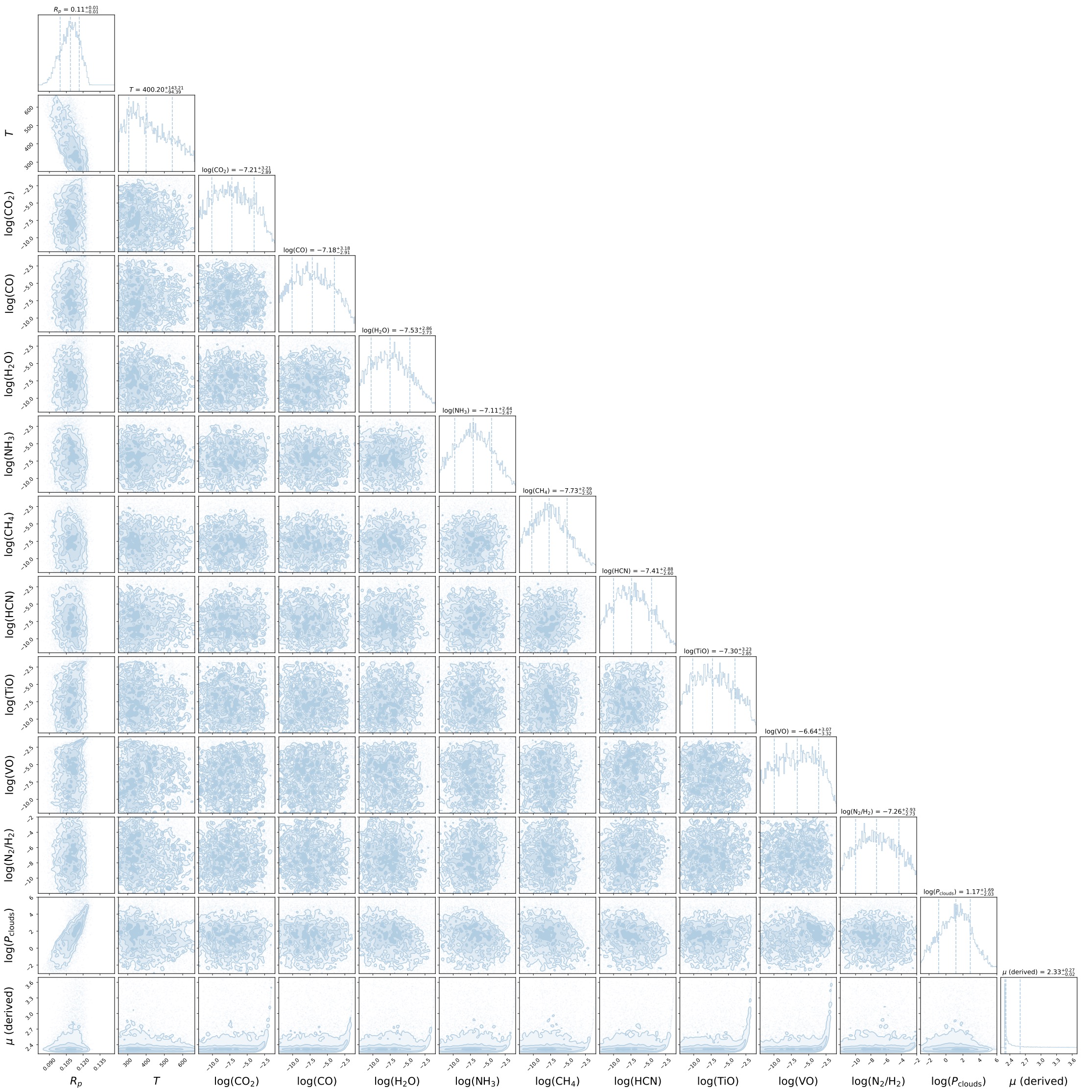}
\caption{The atmospheric retrieval posterior distributions of the cloudy primary atmosphere model for L~98-59~d.}
\label{d_cloudy_primary}
\end{figure*}
\begin{figure*}[t]
\includegraphics[width=1.0\textwidth]{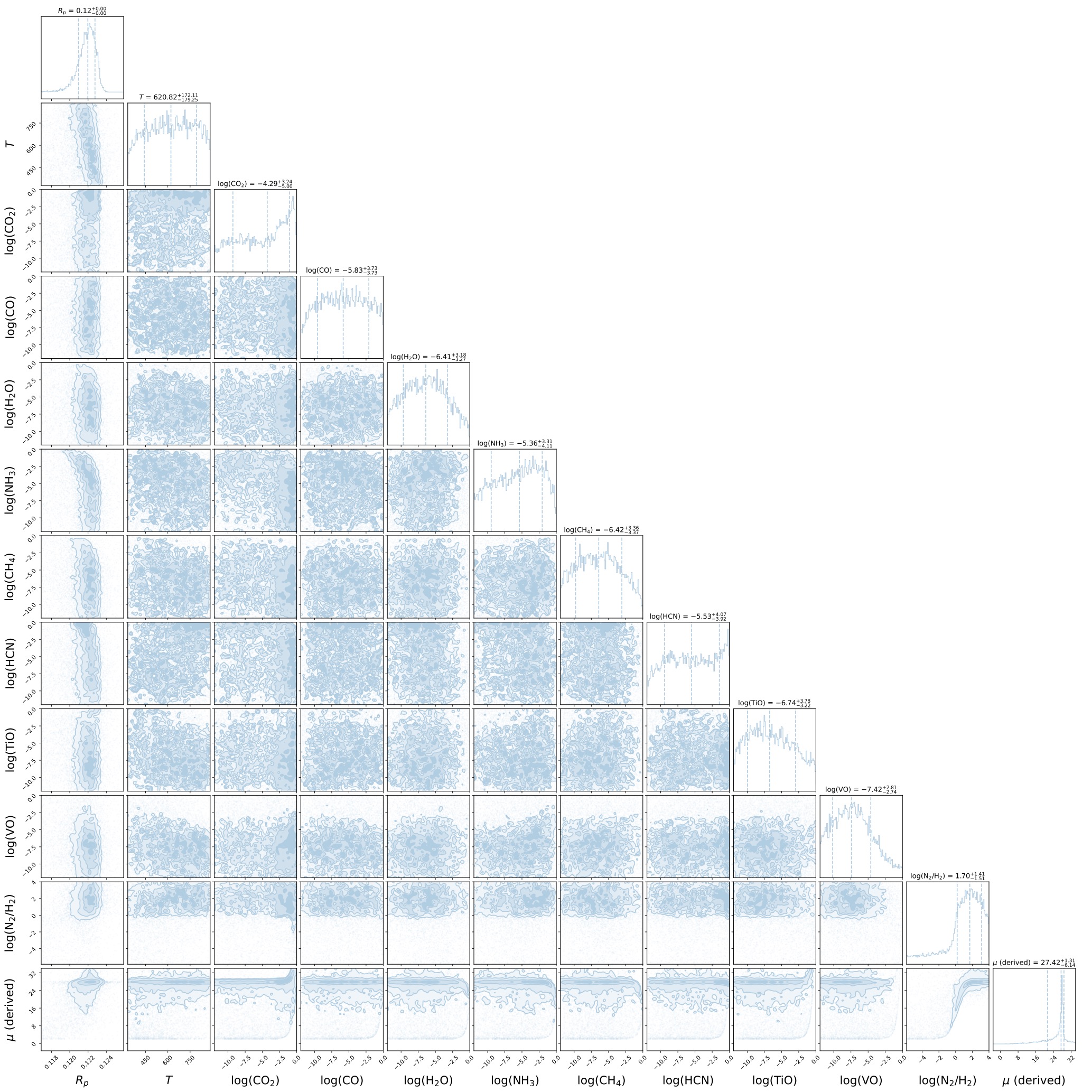}
\caption{The atmospheric retrieval posterior distributions of the clear secondary atmosphere model for L 98-59 c.}
\label{c_clear_secondary}
\end{figure*}
\begin{figure*}[t]
\includegraphics[width=1.0\textwidth]{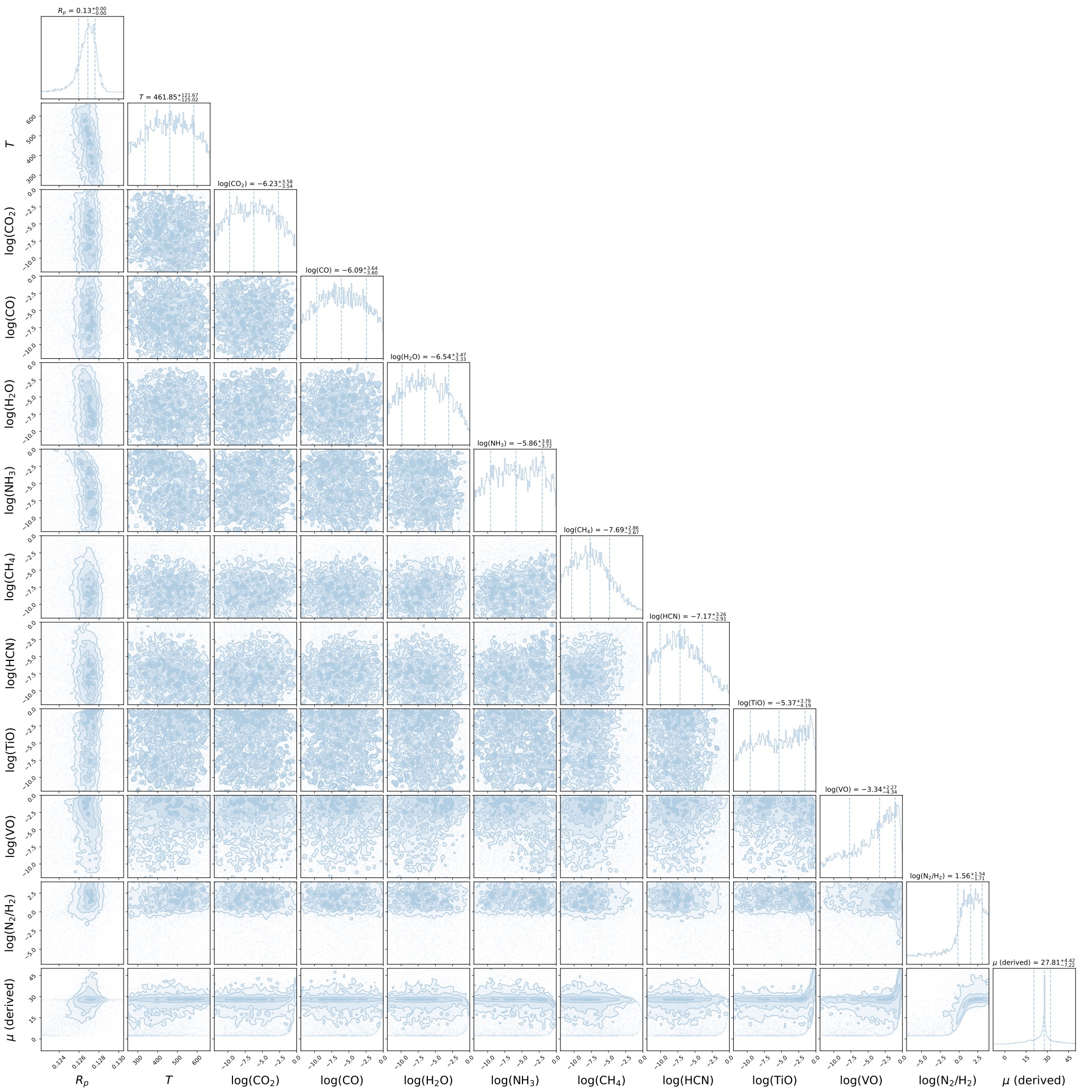}
\caption{The atmospheric retrieval posterior distributions of the clear secondary atmosphere model for L~98-59~d.}
\label{d_clear_secondary}
\end{figure*}
\begin{figure*}[t]
\includegraphics[width=1.0\textwidth]{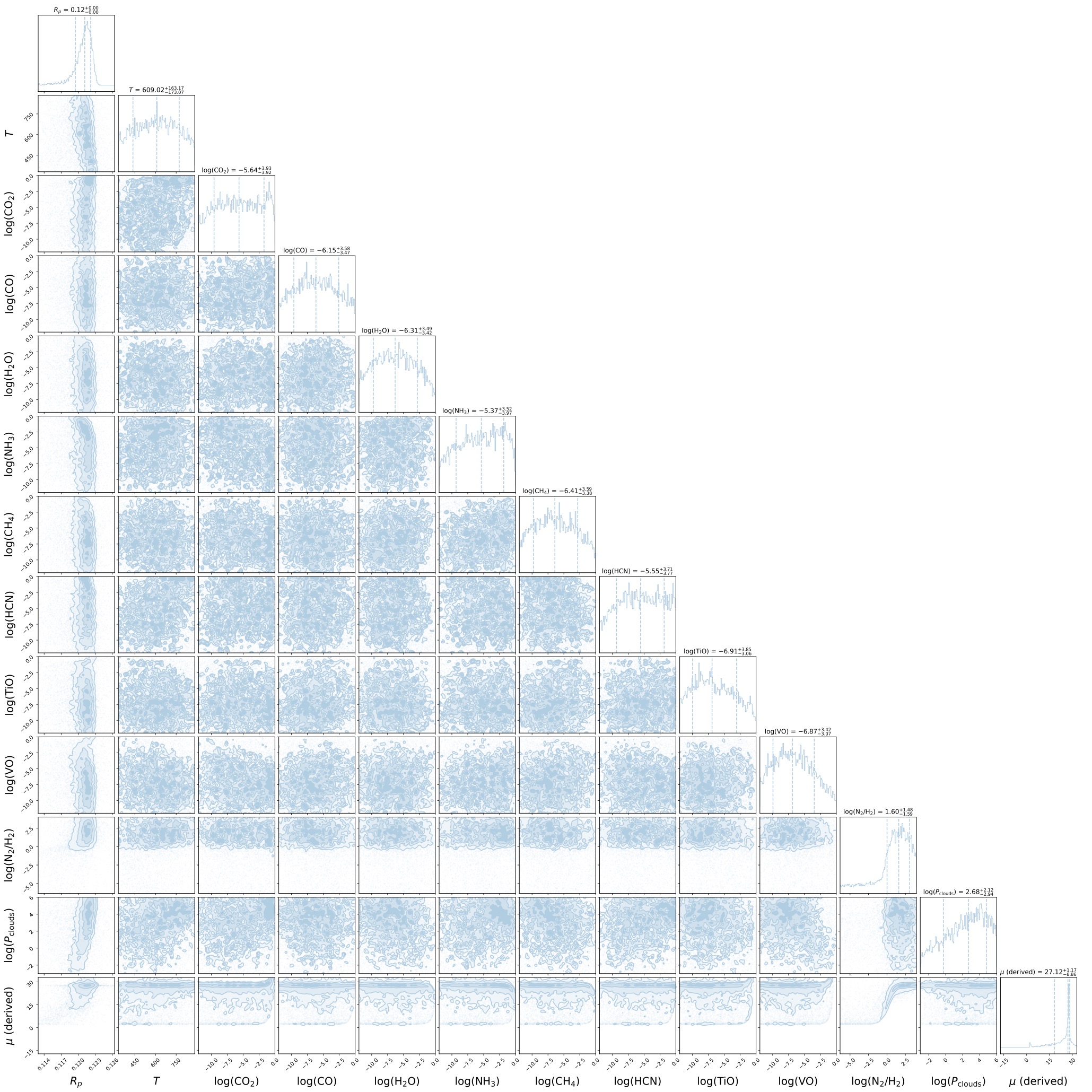}
\caption{The atmospheric retrieval posterior distributions of the best-fit cloudy secondary atmosphere model for L~98-59 c.}
\label{c_cloudy_secondary}
\end{figure*}
\begin{figure*}[t]
\includegraphics[width=1.0\textwidth]{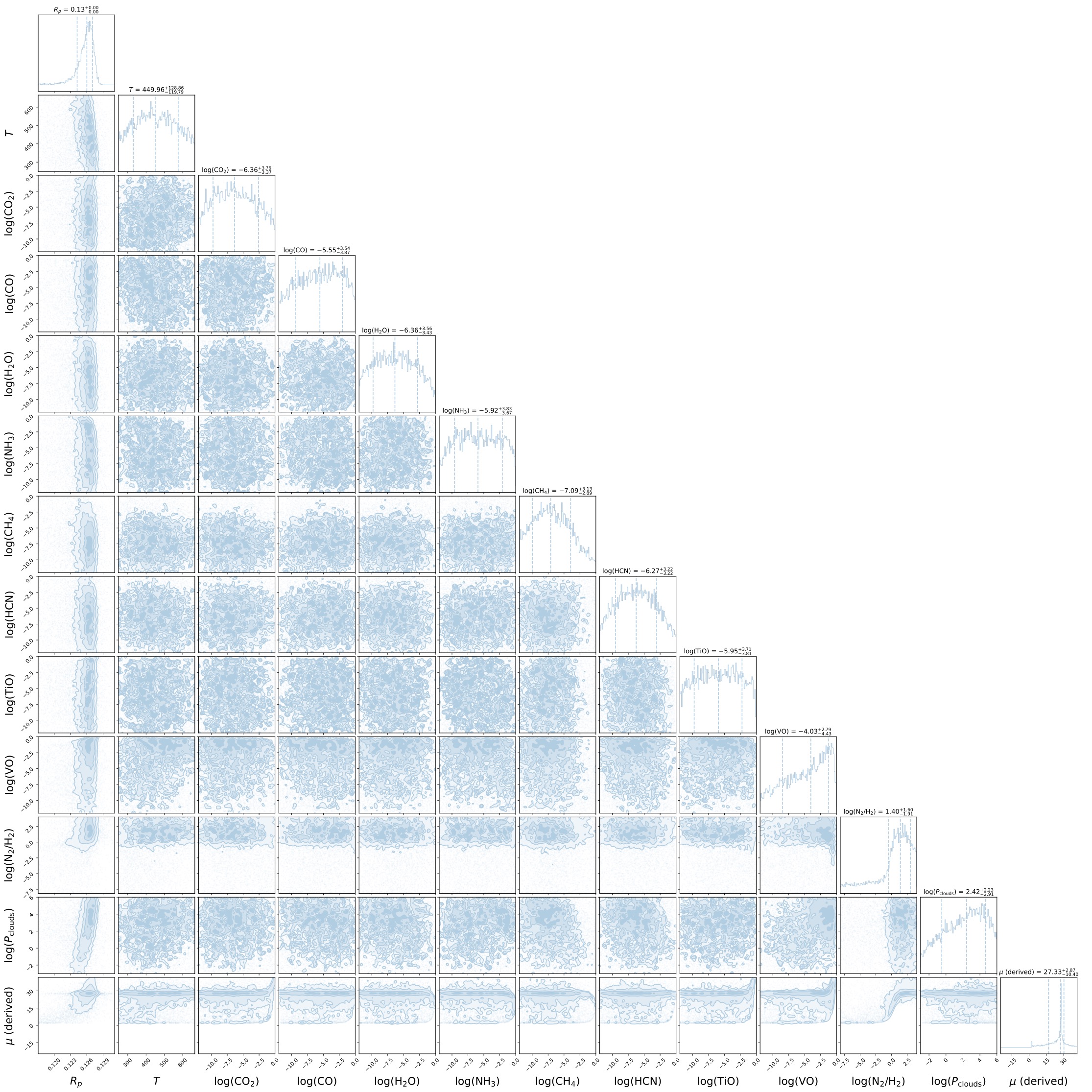}
\caption{The atmospheric retrieval posterior distributions of the best-fit cloudy secondary atmosphere model for L~98-59 d.}
\label{d_cloudy_secondary}
\end{figure*}
\begin{figure*}[t]
\includegraphics[width=1.0\textwidth]{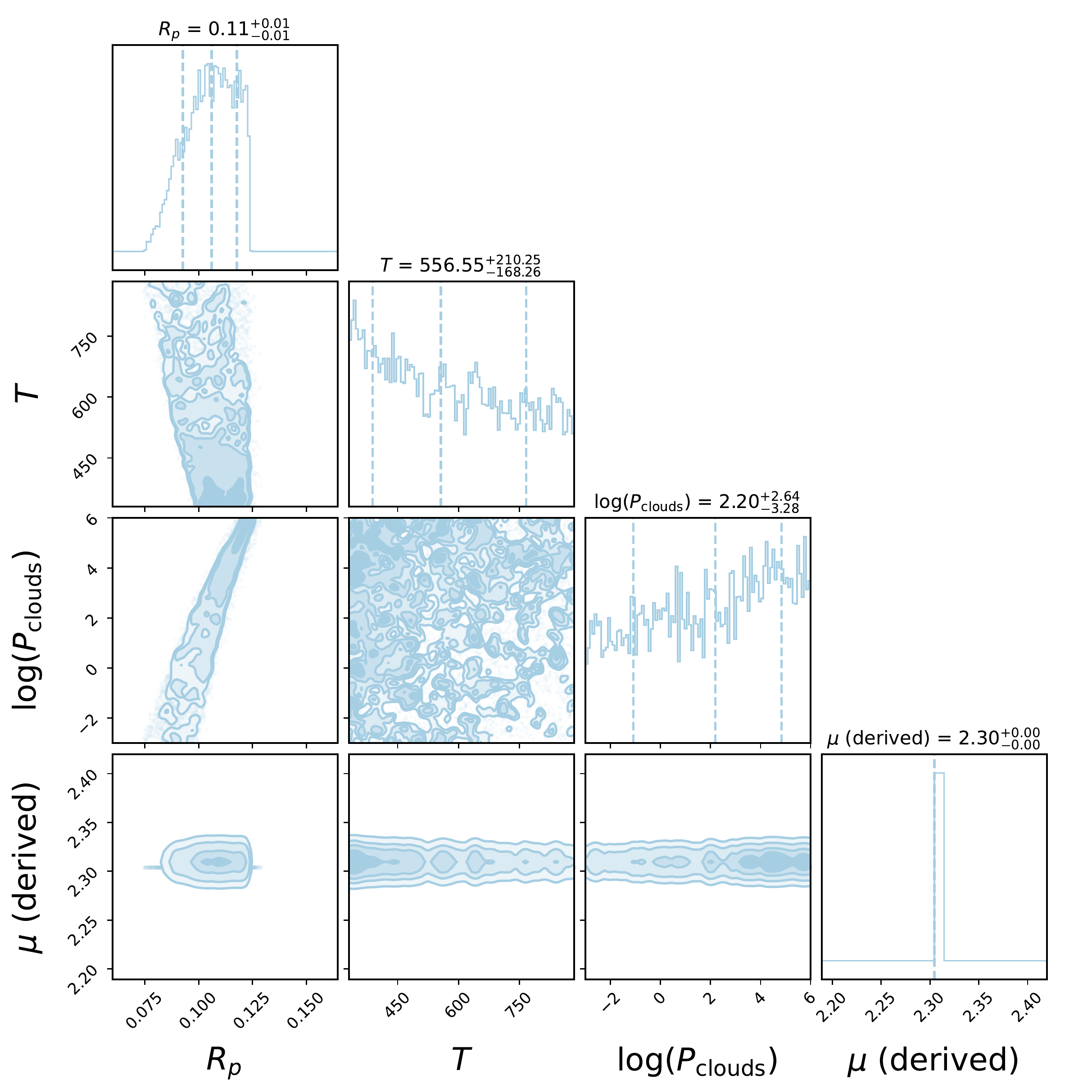}
\caption{The atmospheric retrieval posterior distributions of the pure cloudy atmosphere model for L 98-59 c.}
\label{c_cloudy}
\end{figure*}
\begin{figure*}[t]
\includegraphics[width=1.0\textwidth]{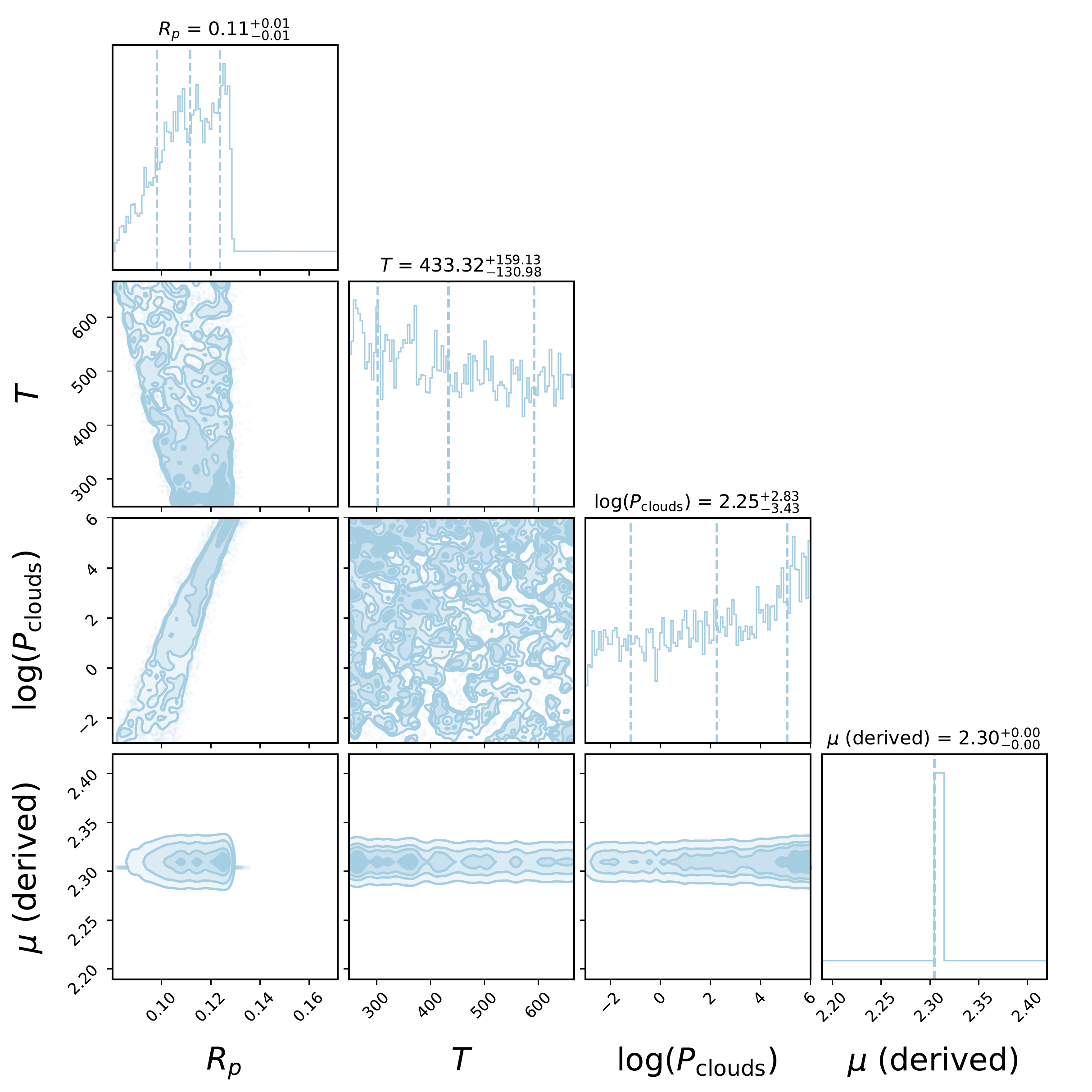}
\caption{The atmospheric retrieval posterior distributions of the pure cloudy atmosphere model for L~98-59~d.}
\label{d_cloudy}
\end{figure*}
\end{document}